\newif\ifec

\ecfalse

\ifec
    \documentclass[11pt,runningheads]{llncs}
\else
    \documentclass[11pt]{article}
    \usepackage[in]{fullpage}
    \usepackage[a4paper,margin=1in]{geometry}
\fi

\usepackage{tyler}

\ifec
    \spnewtheorem{fact}{Fact}{\bfseries}{}
    
\else
    \usepackage{authblk}
    
    \usepackage{amsthm}
    \usepackage{thmtools}
    
    \theoremstyle{definition}
    
    \newtheorem{definition}{Definition}
    
    \newtheorem{lemma}{Lemma}
    
    \newtheorem{corollary}{Corollary}
    \newtheorem{fact}{Fact}

    \newtheorem{remark}{Remark}
    
\fi

\usepackage{thm-restate}
\usepackage{xargs}
\usepackage{enumitem}
\usepackage{yhmath}
\usepackage{ytableau}
\usepackage{nicematrix}
\usepackage{array, multirow, boldline}
\usepackage{float}
\usepackage{tikz-cd}
\usetikzlibrary{quantikz2}
\usetikzlibrary{shapes, arrows.meta, positioning}
\usepackage{dsfont}
\usepackage{bbm}
\usepackage{hyperref}

\hypersetup{
    colorlinks=true,
    linkcolor=blue,
    citecolor=blue,
    filecolor=magenta,      
    urlcolor=cyan,
    pdftitle={Overleaf Example},
    pdfpagemode=FullScreen,
}

\crefname{fact}{Fact}{Facts}
\Crefname{fact}{Fact}{Facts}

\newcommandx{\sym}[2][1=\X,2=]{S_{#1}^{#2}}

\renewcommandx{\P}[3][1=, 2=\big]{\mathop{\mathbb{P}}_{#1} #2 [ #3 #2 ] }
\newcommandx{\E}[3][1=, 2=\big]{\mathop{\mathbb{E}}_{#1} #2 [ #3 #2 ] }

\newcommand{\norm}[1]{\left \lVert #1 \right \rVert}
\newcommand{\out}[2]{\ket{#1}\bra{#2}}

\newcommand{\algo}{\mathcal{A}}
\newcommand{\orcl}{\mathcal{O}}
\newcommand{\BO}[1]{ O \left ( #1 \right ) }
\newcommand{\Samp}{\mathsf{Samp}}
\newcommand{\Verify}{\mathsf{Verify}}
\newcommand{\ans}{\mathsf{ans}}

\newcommand{\high}{\operatorname{high}}
\newcommand{\low}{\operatorname{low}}
\newcommand{\phigh}{\Pi^{\operatorname{high}}}
\newcommand{\im}{\operatorname{Im}}
\newcommand{\Succ}{\operatorname{succ}}
\newcommandx{\Res}[2][1=,2=]{\operatorname{Res}_{#1}^{#2}}
\newcommand{\GL}{\operatorname{GL}}
\newcommand{\U}{\operatorname{U}}

\newcommand{\level}{\operatorname{level}}

\newcommand{\cA}{\mathcal{A}}

\newcommand{\cF}{\mathcal{F}}
\newcommand{\cH}{\mathcal{H}}
\newcommand{\cO}{\mathcal{O}}

\newcommand{\bA}{\mathbf{A}}
\newcommand{\bB}{\mathbf{B}}

\newcommand{\bL}{\mathbf{L}}
\newcommand{\bO}{\mathbf{O}}

\newcommand{\bS}{\mathbf{S}}
\newcommand{\bX}{\mathbf{X}}
\newcommand{\bY}{\mathbf{Y}}
\newcommand{\bW}{\mathbf{W}}

\newcommand{\Span}{\operatorname{span}_{\mathbb{C}}}
\newcommandx{\symd}[1][1=N]{\hat{S}_{#1}}
\newcommandx{\cS}[1][1=\lambda]{\mathcal{S}^{#1}}

\newcommandx{\Pih}[1][1=y]{{\Pi}_{#1}^{\operatorname{high}}}
\newcommandx{\Pil}[1][1=y]{{\Pi}_{#1}^{\operatorname{low}}}
\newcommandx{\PiY}[1][1={\overline{\rho}_{y}}]{\Pi_{\overline{\theta}}^{#1}}
\newcommandx{\PiYu}[1][1={\overline{\rho}_{y}}]{{\Pi}^{\overline{\theta}, #1}}

\iffalse
\newcommand{\siyao}[1]{{\color{blue} (siyao: #1)}}
\newcommand{\akshima}[1]{{\color{cyan} (Akshima: #1)}}

\newcommand{\tyler}[1]{\begingroup\color{magenta} [Tyler: #1] \endgroup}
\else
\newcommand{\siyao}[1]{}
\newcommand{\akshima}[1]{}
\newcommand{\tyler}[1]{}
\fi

\ifec
    \title{Tight Quantum Time-Space Tradeoff for Permutation Inversion}
    
    \author{
        Akshima\inst{1} \and
        Tyler Besselman\inst{1} \and
        Kai-Min Chung\inst{2} \and
        Siyao Guo\inst{1} \and
        Tzu-Yi Yang\inst{2}
    }
    
    \institute{
        NYU Shanghai \and
        Academia Sinica
    }
\else
    \title{Tight Quantum Time-Space Tradeoffs for Permutation Inversion}
\ifec
    \author[1]{Akshima\thanks{akshima@nyu.edu}}
    \author[1]{Tyler Besselman\thanks{tyler.william.b@nyu.edu}}
    \author[2]{Kai-Min Chung\thanks{kmchung@as.edu.tw}}
    \author[1]{Siyao Guo\thanks{siyao.guo@nyu.edu}}
    \author[2]{Tzu-Yi Yang\thanks{yang.ty.math@gmail.com}}
    
    \affil[1]{NYU Shanghai}
    \affil[2]{Academia Sinica}
\else

\author[1]{Akshima}
\author[1]{Tyler Besselman}
\author[2]{Kai-Min Chung}
\author[1]{Siyao Guo}
\author[2]{Tzu-Yi Yang}

\affil[1]{NYU Shanghai}
\affil[1]{\texttt{\{akshima, tyler.william.b, siyao.guo\}@nyu.edu}}
\affil[2]{Academia Sinica}
\affil[2]{\texttt{\{kmchung, dhss6645\}@as.edu.tw}}

\fi

\begin{document}

\maketitle

\begin{abstract}
    In permutation inversion, we are given a permutation $\pi:[N]\rightarrow[N]$, and want to prepare some advice of size $S$, such that we can efficiently invert any image in time $T$.  This is a fundamental cryptographic problem with profound connections to communication complexity and circuit lower bounds. 

    In the classical setting, a tight $ST=\tilde{\Theta}(N)$ bound has been established since the seminal work of Hellman (1980) and Yao (1990).  In the quantum setting, a lower bound of $ST^2=\tilde{\Omega}(N)$ is proved by Nayebi, Aaronson, Belovs, and Trevisan (2015) against classical advice, and by Hhan, Xagawa and Yamakawa (2019) against quantum advice. It left open an intriguing possibility that Grover’s search can be sped up to time $\Tilde{O}(\sqrt{N/S})$.

    In this work, we prove an $ST+T^2=\Omega(N)$ lower bound for permutation inversion with even quantum advice. This bound matches the best known attacks and shows that Grover's search and the classical Hellman's algorithm cannot be further sped up. 

    Our proof combines recent techniques by Liu (2023) and by Rosmanis (2022). Specifically, we first reduce the permutation inversion problem against quantum advice to a variant by Liu's technique, then we analyze this variant via representation theory inspired by Rosmanis (2022).
\end{abstract}

\section{Introduction}

Given an unknown permutation $\pi$ from $S_N$ (the set of all permutations from $[N]$ to $[N]$) as an oracle, and an arbitrary image $y\in[N]$, the problem of permutation inversion aims to efficiently output $\pi^{-1}(y)$. This is a fundamental cryptographic problem with profound connections to communication complexity and circuit lower bounds\footnote{Corrigan-Gibbs and Kogan~\cite{corrigan2019function} showed that permutation inversion algorithms are useful in designing new communication protocols for a well-studied problem in communication complexity.  Specifically, new permutation inversion algorithms yield new protocols for multiparty pointer jumping problem, a problem with significance to $\mathrm{ACC}^0$ circuit lower bounds. We refer interested readers to~\cite{corrigan2019function} for the details.}.

Throughout this paper, we focus on the query complexity, i.e., the number of queries to $\pi$. Because the number of made queries  trivially lower bounds the required time, and our main goal is proving lower bounds on the required time (and space), we use the query complexity as the main time efficiency measure for the simplicity of the presentation.

In the classical setting, the optimal bounds for the permutation inversion problem is well understood since 1990. Without knowing any information about $\pi$, it is straightforward to see that $\Theta(N)$ queries are necessary and sufficient.
However, an interesting situation arises when precomputed information about $\pi$ is allowed.
In the seminal work~\cite{hellman2003cryptanalytic}, Hellman gave a classical algorithm that inverts any image with $T = \tilde{O}(N/S)$ queries using an $S$-bit precomputed advice string. (The notation $\tilde{O}(\cdot)$ and $\tilde{\Omega}(\cdot)$
hide lower order factors that are polynomial in $\log N$).  In 1990, Yao~\cite{Yao90} proved an $ST=\tilde{\Omega}(N)$ lower bound showing that this algorithm cannot be further improved. 

In the quantum setting, our understanding is much less satisfying. Without any preprocessing, Grover's algorithm~\cite{grover1996fast} can solve the permutation inversion problem in $O(\sqrt{N})$ quantum queries (thus beating the classical $\Omega(N)$ bound), and is known to be asymptotically optimal (\emph{e.g.} ~\cite{BBBV97,Amb02,nay10,nayebi2014quantum,rosmanis:2022}).
In 2015, Nayebi, Aaronson, Belovs, and Trevisan \cite{nayebi2014quantum} first studied the preprocessing setting, and proved a lower bound of $ST^2 = \tilde{\Omega}(N)$ against classical advice. Later, Hhan, Xagawa and Yamakawa~\cite{hhan2019quantum} extended this lower bound to the quantum advice setting. It left open an intriguing possibility that Grover’s search can be sped up to $\tilde{O}(\sqrt{N/S})$ online queries.
This raises the following question:

\begin{center}\begin{minipage}{0.8\linewidth}
     Can Grover's search and Hellman's algorithm be combined to speed up for the permutation inversion problem? 
\end{minipage}\end{center}

 Our main theorem below gives a negative answer, showing that the optimal quantum time–space tradeoff for permutation inversion matches known algorithms.

\begin{restatable}{theorem}{ThmMain}\label{ThmMain}
Let $\algo = (\algo_1, \algo_2)$ be an Auxiliary-Input algorithm for permutation inversion problem consisting of two stages:
\begin{enumerate}
    \item $\algo_1$ is given unbounded access to a permutation $\pi : [N] \to [N]$ and outputs an $S$-qubit advice state $\algo_1(\pi)$.
    \item Given the state $\algo_1(\pi)$ and a challenge point $y \in [N]$, $\algo_2^\pi$ makes at most $T$ quantum queries to $\pi$ and outputs a point $x \in [N]$.
\end{enumerate}
Then, the success probability satisfies
\[
    \P[\pi, y][\Big]{
        \pi \left ( \algo_{2}^{\pi} \big ( \algo_{1}(\pi), y \big ) \right ) = y
    } = \BO{ \frac{ST}{N} + \frac{T^{2}}{N} }
\]
where $\pi \leftarrow S_N$ and $y \leftarrow [N]$ are sampled uniformly.
\end{restatable}

Our bound implies that for a quantum preprocessing algorithm to invert any image of an arbitrary permutation, it must satisfy $ST+T^{2} = \Omega(N)$ even for the case of quantum advice. 
This matches the best known algorithms (up to polylog factors): Grover’s search when $S \leq T$ and Hellman’s classical method when $S > T$.

Moreover, our bound also provides an optimal security upper bound for any quantum preprocessing algorithm to invert a random permutation on a random given image.  It implies that  the $\Omega(ST/N)$ advantage of Hellman's algorithm and the $\Omega(T^2/N)$ advantage of Grover's search cannot be further improved. 

We remark that the same security upper bound $O(ST/N+T^2/N)$ has been proved for the case of inverting a random function by Chung, Guo, Liu and Qian~\cite{kmsy:2020} against classical advice, and Liu~\cite{qipeng:2023} against quantum advice. Interestingly, although it is a major open problem to close the gap between the above security bound and known algorithms for function inversion (see~\cite{corrigan2019function}), the same bound suffices to establish the optimal time-space tradeoffs for the permutation inversion problem.  However, their main techniques are limited to random functions, and the best known security upper bound for the permutation case prior to our work remains $O(ST^2/N)$ by Nayebi et al.~\cite{nayebi2014quantum} against classical advice, and $O((ST^2/N)^{1/3})$ by Hhan et al.~\cite{hhan2019quantum} against quantum advice.  


In particular, both works used the compressed oracle framework (see \cite{zhandry2019} for details) to prove their result.  It is worth noting that the compressed oracle framework, a quantum analogue to the classical lazy sampling technique, does not apply to permutations.

In fact, for random permutations, no framework comparable to the compressed oracle is known so far despite several attempts to create one (\cite{Unruh2021,rosmanis:2022,Unruh2023,majenz2024permutation,Carolan2025}).
This is possibly what prevented \cite{kmsy:2020} and \cite{qipeng:2023} from extending their results to the permutation setting, leaving the gap between the best known attack and the lower bound in \cite{nayebi2014quantum} open. In this work, we resolve this open problem.

Our proof of \cref{ThmMain} expands on techniques from a recent work by Rosmanis \cite{rosmanis:2022}. Rosmanis proposed a method for analyzing quantum algorithms solving the permutation inversion (without pre-computation) using techniques from the representation theory. First, we reduce the permutation inversion problem with preprocessing to the ``bit-fixing'' model (we give a formal definition of the model in the technical overview). A recent work of Liu \cite{qipeng:2023} showed such a reduction for random functions, and we extend these reduction techniques to the permutation setting. We then use our extension of Rosmanis' method to analyze the permutation inversion problem in the ``bit-fixing'' model.

We refer the reader to the next subsection for a detailed overview of our proof.

\subsection{Technical Overview}
We now provide an overview of the proof of our main theorem (\cref{ThmMain}), highlighting the technical challenges we address compared to prior work.

\paragraph{Reduction to the Bit-Fixing Model.} 

As a first step toward resolving this, we reduce permutation inversion with preprocessing to the \emph{bit-fixing model} in a similar way as Liu \cite{qipeng:2023}.
Since the proofs of \cite{qipeng:2023} carry over essentially unchanged from the case of inverting random functions, we only provide a brief argument in \cref{AppBitFix} for completeness.

A quantum algorithm for permutation inversion in the \emph{$P$-bit fixing model} with $T$-quantum queries (or $(P,T)$-algorithm for short) is a two stage algorithm in which
\begin{itemize}
    \item[--] \textbf{Offline phase.} A uniform random permutation $\pi \leftarrow \sym[N]$ is sampled. The algorithm makes $P$ quantum queries to $\pi$ before outputting a bit $b$. This stage repeats until $b = 0$.
    \item[--] \textbf{Online phase.} A uniform challenge $y \in [N]$ is sampled. Continuing with the inner register of the offline phase, the algorithm is given $y$, makes $T$ further queries to $\pi$, and outputs an answer $x$.
\end{itemize}
The algorithm succeeds if $\pi(x) = y$.

Intuitively, the first phase allows the algorithm to bias the distribution of $\pi$, granting partial information about $\pi$ (conditioned on $b = 0$) before the challenge is revealed.
The following reduction then follows from the same techniques as \cite{qipeng:2023}:

\begin{restatable}[Auxiliary-Input to Bit-Fixing]{lemma}{Reduction}\label{LemReduction}
Let $\algo$ be an algorithm for permutation inversion problem with a preprocessed $S$-qubit advice and $T$ quantum queries. Then, for $P = S(T + 1)$, there exists an $P$-bit-fixing algorithm $\mathcal{B}$ making $T$-quantum queries, inverting a uniformly sampled $y \in [N]$ such that
\[
    \P[\pi, y]{
        \pi \left ( \mathcal{B}^{\pi}(y) \right ) = y
    } \ge \frac{1}{2} \P[\pi, y]{
        \pi \left ( \algo_{2}^{\pi}(\algo_{1}(\pi), y) \right ) = y
    }.
\]
\end{restatable}

Our main theorem follows immediately from this reduction and the following bound:
\begin{restatable}{lemma}{Bitfixing}\label{LemBitfixing}
    Any $P$-bit-fixing algorithm $\algo$ with $T$-quantum queries that inverts a uniformly sampled challenge $y \in [N]$ has the success probability \[
        \P[\pi, y]{
            \pi \left ( \algo^{\pi}(y) \right ) = y
        } = O \left ( \frac{P}{N} + \frac{T^{2}}{N} \right ).
    \]
\end{restatable}

At a high level, the $T^{2} / N$ term above reflects the quadratic speedup achieved by Grover search \emph{after} receiving the challenge, while the $P / N$ term captures the fact that quantum queries made \emph{before} the challenge provide essentially no advantage.
This is the main point we argue in our proof. We remark that \cite{kmsy:2020} (see their Lemma 1.5) proved the same statement for the case of inverting random functions, but their techniques (i.e., compressed oracle framework~\cite{zhandry2019}) are limited to random functions. We view Lemma~\ref{LemBitfixing} as our main technical contribution. We first reduce \cref{LemBitfixing} to a single statement,
    called the ``average bound'' (\cref{SecFramework}),
    and then prove this bound using the representation theory of symmetric groups
    (\cref{SecAvgbound}).


\paragraph{Strategy for Achieving Quantum Bit-Fixing Upper Bound.}
The strategy for proving \cref{LemBitfixing} is as follows.
We adopt a purified view of the random permutation model.
An algorithm $\cA$ for the inversion problem in the bit-fixing model makes a quantum query by interacting with the oracle register $\bO$ via the unitary
\[
    \orcl : \ket{\pi}_{\bO} \ket{x}_{\bX} \ket{y}_{\bY}
        \mapsto \ket{\pi}_{\bO} \ket{x}_{\bX} \ket{y + \pi(x)}_{\bY}
\]
where the oracle register $\bO$ is initialized to the uniform superposition over all permutations and measured at the end of computation.
Let $\ket{\psi_{k}}$ be the joint state of the oracle register and algorithm state after $k$ queries,
    and let $p^{y}_{k}$ denote the algorithm's success probability on a fixed challenge $y \in [N]$.

In order to analyze the success probability of bit fixing algorithms, our goal is to design a good projection $\Pih$ (which will be defined later) on the oracle register that approximately characterizes whether or not the challenge point $y$ has been inverted or not.
We call the states in the image of $\Pih$ a ``database'' inverting $y$.
Then, we approximate the success probability $p^{y}_{k}$ by the value
\[
    \tilde{p}^{y}_{k} \coloneq \norm{
        \left ( \Pih \otimes I_{\bA} \right ) \ket{\psi_{k}}
    }^{2}
\]
which effectively measures how much entanglement is shared between the algorithm state and a database state including an inversion of $y$.

Then, we can prove the following relations showing first that this approximation is ``close enough'' to the true success probability,
    and then that each query made by the algorithm cannot improve this approximate success probability too much.
These two points follow from the main theorem of \cite{rosmanis:2022}.
Formally, for any $k \ll N$ we have:
\begin{enumerate}[label=(\roman*)]
    \item $\sqrt{p^{y}_{k}} = \sqrt{\tilde{p}^{y}_{k}} + O(1 / \sqrt{N})$
    \item $\sqrt{\tilde{p}^{y}_{k + 1}} = \sqrt{\tilde{p}^{y}_{k}} + O(1 / \sqrt{N})$
\end{enumerate}

From these two results, we can obtain a tight bound on permutation inversion without advice matching Grover search.
In order to prove the desired bound in the $P$-bit-fixing setting, however,
we have to further show that the success probability of an algorithm after receiving the challenge point $y$ but before making any online queries is sufficiently small.
In particular, we require the following statement, which we call the \emph{average bound} (see \cref{LemAvg}):
\[
\E[y]{ \tilde{p}_{k}^{y} }
        \coloneq \frac{1}{N} \sum_{y \in [N]} \norm{ (\Pih\otimes I_{\bA}) \ket{\psi_k} }^{2}
        = \BO{ \frac{k}{N} }
\]
for any $k \ll N$, provided that $\ket{\psi_k}$ does not depend on $y$.
This means that after $k$ challenge-independent queries,
    the expected success probability over all possible challenges is at most $k / N$.
In other words, the speedup of Grover search is achieved only by focusing on a specific challenge point.
Note that (i) and (ii) above imply only that $\E[y]{ \tilde{p}^{y}_{k} } = \BO{ k^{2} / N }$. As so, the average bound is a non-trivial extension of \cite{rosmanis:2022}'s results.

Finally, we show that for a $P$-bit fixing algorithm making $P$ challenge-independent queries followed by $T$ adaptive queries after receiving a challenge point $y$,
    the overall success probability is bounded by
\[
    \E[y]{ p^{y}_{P + T} }
    \le 2 \E[y][\bigg]{ \tilde{p}^{y}_{P} + c^2 \frac{T^{2}}{N} }
    =   2 \E[y][\Big]{ \tilde{p}^{y}_{P} } + 2c^2 \frac{T^2}{N}
        =  \BO{ \frac{P}{N} + \frac{T^2}{N} }
\]
where the first inequality uses (i) and (ii) to prove $\sqrt{ p^{y}_{P + T} } \le \sqrt{ \tilde{p}^{y}_{P} } + c T / N$,
    and the last equality follows from the average bound above.

\paragraph{Comparison to Compressed Oracle \cite{zhandry2019}.} When inverting random functions, one can rely on the \emph{compressed oracle} framework introduced by \cite{zhandry2019} and follow a similar strategy above to achieve an upper bound for function inversion problem.
This framework provides a convenient way to record oracle outputs $x \in [N]$ independently and has become a robust tool in query complexity of problems with random functions.
By extending the oracle register with special symbols $\perp$,
    it is possible to represent explicit databases as orthogonal vectors $\ket{D}$ describing the partial truth table of the oracle.
Precisely, for each input $x\in [N]$, $D$ assigns an output $y\in [N]$ or $\perp$ indicating that $x$ has not been queried yet.
The projection $\Pih$ is in this case defined as $\sum_{D:~y\in D}\ket{D}\bra{D}$ where $y \in D$ means that $y$ is assigned by some $x\in [N]$ \emph{i.e.} at least one preimage of $y$ has been found. Notice that the way we define $\Pih$ is based on the orthogonality of databases $\ket{D}$, which is crucial for many other analysis using the compressed oracle.
In this setting, the equalities (i) and (ii) were first established for random oracles in \cite[Theorem 1]{zhandry2019}.
To derive the time–space tradeoff, an equivalent to our average bound was proved implicitly in \cite[Proposition 5.7]{kmsy:2020} using the compressed oracle as a special case of the bit-fixing model.
A general formalism of the bit-fixing model and its upper bound was given later in \cite{GLLZ2021}. 

\paragraph{Difficulties of Constructing Explicit Databases for Permutation} For random permutations, no comparable framework to the compressed oracle is known so far.
Several attempts have been made to mimic it, including \cite{Unruh2021,rosmanis:2022,Unruh2023,majenz2024permutation,Carolan2025}.
However, as emphasized in \cite{majenz2024permutation}, one cannot expect a single framework to inherit all the advantages of the compressed oracle; some tradeoff is unavoidable.
For instance, the variables recording outputs of different points cannot be made independent without sacrificing  tightness of possible bounds.
In this work, we adopt the method proposed by Rosmanis \cite{rosmanis:2022},
    which in some sense captures database-like states (\emph{i.e.} states in $\bO$) using the \emph{irreducible representations} (``irreps'' for short) of the symmetric group $\sym[N]$.
This approach does not fully capture the exact pointwise mapping, but it encodes key structural information:
    irreps perfectly track the number of queries made (see \cref{Lem1}) and,
    in a more subtle way,
    allow us to determine whether certain points are assigned by previous query outputs
    with small error (the inequality (i) or \cref{Lem2}).
Moreover, as those irreps recording information are orthogonal to each other, we can still inherit a similar methodology to that of the compressed oracle technique for our analysis.

\paragraph{Framework of \cite{rosmanis:2022} via Representation Theory.}
First, we recall some ideas from Rosmanis' previous work using \emph{representation theory} and refer the reader to \cref{AppRep} for the necessary mathematical background.
At a high level, representation theory enables us to exploit the algebraic structure—the action of the symmetric groups—on the oracle’s Hilbert space. These structures are well understood combinatorially via Young diagrams and our analysis of the oracle’s database is essentially based on the manipulation of such diagrams.

In our setting, the oracle register $\bO$ is initialized to the uniform superposition state in the Hilbert space
$\cF = \Span \big \{ \ket{\pi} \mid \pi \in \sym[N] \big \}$.
This Hilbert space can be regarded as an $\sym[N] \times \sym[N]$-representation, which means that $\cF$ is endowed with an additional group action of $\sym[N] \times \sym[N]$ where the left and right copies of $\sym[N]$ act on the domain and range of $\pi$, respectively (see (\ref{EqRegular}) for the precise description).
From standard results of representation theory, it follows that $\cF$ decomposes to a direct sum of tensor products $\lambda \otimes \lambda$ over all irreducible representations (irreps) $\lambda$ of $\sym[N]$. Moreover, every $\sym[N]$-irrep corresponds bijectively to a Young diagram of size $N$, that is, to a partition of $N$ into non-increasing integers (see the figure below). For convenience, we will interchangeably use the same symbol $\lambda$ to denote both the irrep and its corresponding Young diagram, and accordingly index the subspaces $\cH_{\lambda}:=\lambda \otimes \lambda$ in $\cF$ by the Young diagram $\lambda$ that fully captures the information.

Interestingly, the number of queries made is captured by a combinatorial property of the Young diagram:
    after $k$ queries, the algorithm's state is entangled with states in $\cH_{\lambda}$ (database-like states) indexed by diagrams $\lambda$ having at most $k$ boxes below the first row (see Lemma \ref{Lem1} and Lemma \ref{PropDecompA}).
For example, when $N = 5$, the possible database-like states after the $i$-th query are indexed by Young diagrams with level $\le i$:
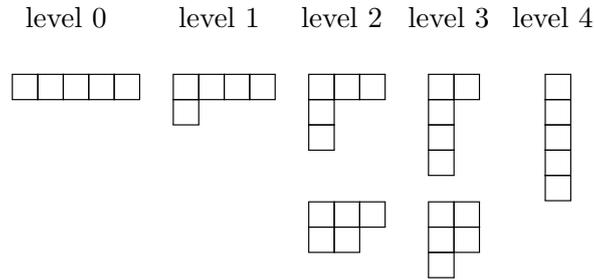
\begin{center}
    \begin{tabular}{@{\hspace{0.3cm}}c@{\hspace{0.3cm}}
                c@{\hspace{0.3cm}}
                c@{\hspace{0.3cm}}
                c@{\hspace{0.3cm}}
                c@{\hspace{0.3cm}}}
        level $0$ & level $1$ & level $2$ & level $3$ & level $4$ \\
        &&&&\\
        \ytableausetup{smalltableaux}
        \begin{ytableau}
            {} & {} & {} & {} & {} 
        \end{ytableau} &
        \ytableausetup{smalltableaux}
        \begin{ytableau}
            {} & {} & {} & {} \\
            {}
        \end{ytableau} &
        \ytableausetup{smalltableaux}
        \begin{ytableau}
            {} & {} & {} \\
            {} \\
            {}
        \end{ytableau} &
        \ytableausetup{smalltableaux}
        \begin{ytableau}
            {} & {} \\
            {} \\
            {} \\
            {}
        \end{ytableau} &
        \ytableausetup{smalltableaux}
        \begin{ytableau}
            {} \\
            {} \\
            {} \\
            {} \\
            {}
        \end{ytableau} \\
        &&
        \ytableausetup{smalltableaux}
        \begin{ytableau}
            {} & {} & {} \\
            {} & {}
        \end{ytableau} &
        \ytableausetup{smalltableaux}
        \begin{ytableau}
            {} & {} \\
            {} & {} \\
            {}
        \end{ytableau} & \\
    \end{tabular}
\end{center}
A caveat is that these Young diagrams index subspaces, not individual vectors, which is quite different from the compressed oracle.

To check whether a point $y\in [N]$ appears in the range of the database (\emph{i.e.}, whether some query output equals $y$),
    we use the \emph{branching rule} to further decompose the right tensor component of $\lambda \otimes \lambda$:
\[ \lambda \otimes \lambda = \bigoplus_{\mu \prec \lambda} \lambda \otimes \mu_{y} \]
where $\mu \prec \lambda$ means $\mu$ is obtained by removing one box from $\lambda$ and $\mu_{y}$ is the corresponding irrep of $S_{[N] \setminus \{y\}}$ (the subgroup of $S_N$ consisting of permutations fixing $y$ which is isomorphic to $S_{N-1}$).
Rosmanis \cite{rosmanis:2022} shows that the subspaces $\lambda \otimes \mu_{y}$ containing $y$
    are precisely those where the removed box is not the last box of the first row, up to error $\BO{ 1 / \sqrt{N} }$.
We may also index such subspaces $\lambda \otimes \mu_{y}$ by marking $y$ in $\lambda$ the box that $\mu$ removes.
For example, the following decomposition illustrates this idea:
\begin{equation*}
    \vcenter{
        \hbox{
            \begin{ytableau}
                \ & \ & \ & \  \\
                \ & \  \\
                \
            \end{ytableau}
        }
    } = \vcenter{
        \hbox{
            \begin{ytableau}
                \ & \ & \ & \  \\
                \ & \  \\
                y
            \end{ytableau}
        }
    } \oplus \vcenter{
        \hbox{
            \begin{ytableau}
                \ & \ & \ & \  \\
                \ & y  \\
                \
            \end{ytableau}
        }
    } \oplus \vcenter{
        \hbox{
            \begin{ytableau}
                \ & \ & \ & y  \\
                \ & \  \\
                \
            \end{ytableau}
        }
    }
\end{equation*}
Here, the first two subspaces correspond to cases where $y$ is present in the range, while the last subspace does not. Finally, we define $\Pih$ as the projection onto the direct sum of all $\lambda\otimes\mu_y$ with $\mu$ obtained by removing a box other than the last box in the first row.

\paragraph{Trick for Average Bound.} The main difficulty in proving our average bound in this setting is that although the subspaces $\lambda \otimes \lambda$ are mutually orthogonal,
    the subspaces $\lambda \otimes \mu_{y}$ defining $\Pih$ for different $y$ are not.
For example, the following subspaces indexed by
\begin{equation*}
    \ytableausetup{smalltableaux}
    \begin{ytableau}
        {} & {} & {} \\
        {} & {1}
    \end{ytableau}
    \qquad
    \ytableausetup{smalltableaux}
    \begin{ytableau}
        {} & {} & {} \\
        {} & {2}
    \end{ytableau}
    \qquad
    \ytableausetup{smalltableaux}
    \begin{ytableau}
        {} & {} & {3} \\
        {} & {}
    \end{ytableau}
\end{equation*}
are neither identical nor orthogonal.
Thus, instead of estimating the individual quantities $\tilde{p}_{k}^{y} = \norm{ \Pih \ket{\psi_{k}} }^{2}$ for each $y$ separately,
    we define the operator
\[ M \coloneq \sum_{y \in [N]} \Pih \]
and exploit the observation that
\[
    \sum_{y \in [N]} \tilde{p}^{y}_{k} = \sum_{y \in [N]} \bra{\psi_{k}} \Pih\otimes I_{\bA} \ket{\psi_{k}}
        = \bra{\psi_{k}} M\otimes I_{\bA} \ket{\psi_{k}}
        = \norm{ \sqrt{M}\otimes I_{\bA} \ket{\psi_{k}} }^{2}.
\]

As a sum over $y \in [N]$, $M$ enjoys a certain symmetry (\cref{ChangeCoord}): it is a homomorphism of $\sym[N] \times \sym[N]$ representations.
By Schur’s lemma, any such homomorphism acts as a scalar on each irrep and vanishes between distinct irreps.
Consequently, $M$ is block-diagonal in a basis compatible with all subspaces $\lambda \otimes \lambda$ on which $M$ acts constantly. Precisely,
$$M=\sum_{\lambda}e_{\lambda}\Pi_{\lambda}$$
where $\Pi_{\lambda}$ is the orthogonal projection on the subspace $\lambda\otimes\lambda$ and $e_{\lambda}$ is the corresponding eigenvalue.

For each Young diagram $\lambda$ of size $N$, we let $\{\ket{v_{\lambda,i}}\}_{i\in [d^2_{\lambda}]}$ be an orthonormal basis of the $d^2_{\lambda}$-dimensional subspace $\lambda\otimes \lambda$ where $d_{\lambda}$ is the dimension of the irrep $\lambda$. As discussed above, the overall state $\ket{\psi_{k}}$ after $k$-queries is of the form
$$\sum_{\substack{\lambda:~\level(\lambda)\le k\\ i\in [d_{\lambda}^2]}}a_{\lambda,i}\ket{v_{\lambda,i}}_{\bO}\otimes \ket{z_{\lambda,i}}\qquad\text{with}\qquad \sum|a_{\lambda,i}|^2=1$$
that the algorithm's state is entangled with states in subspaces $\lambda \otimes \lambda$ with $\level(\lambda)\le k$. Hence
$$\norm{ \sqrt{M}\otimes I_{\bA} \ket{\psi_{k}} }^{2}=\sum_{\substack{\lambda:~\level(\lambda)\le k\\ i\in [d_{\lambda}^2]}} |a_{\lambda,i}|^2\norm{ \sqrt{M}\ket{v_{\lambda,i}} }^{2}\le \max_{\operatorname{level}(\lambda) \le k} e_{\lambda}$$
and we suffice to compute each eigenvalue $e_{\lambda}$.

\paragraph{Explicit Computation via Combinatorics of Young Diagram.} For any Young diagram $\lambda$ of size $N$, the trace of the block of $M$ on the subspace $\lambda \otimes \lambda$ (of dimension $d_\lambda^2$) is $e_\lambda \cdot d_\lambda^2$. The idea here is to re-express $\Pih$ in terms of the irreps $\overline{\theta} \otimes \overline{\rho}_y$ and compute the trace of this block again. Together with the relation
\[ d_{\lambda} = \sum_{\mu \prec \lambda} d_{\mu} \]
given by the branching rule, one can derive a formula for the eigenvalue
\[ e_{\lambda} = N \left ( 1 - \frac{d_{\lambda'_{*}}}{d_\lambda} \right ) \]
or $e_{\lambda} = N$ if $\lambda'_{*}$ does not exist.
Here $\lambda'_*$ denotes the Young diagram obtained by removing the last box of the first row of $\lambda$,
    and $d_{\lambda'_*}$ is the dimension of the corresponding irrep of $\sym[N \setminus\{y\}]$ for any $y$.
Since the dimensions of irreps can be computed combinatorially via Young diagrams,
    this gives an explicit way to evaluate $e_{\lambda}$.

For example, when $N = 3$ there are three irreps corresponding to
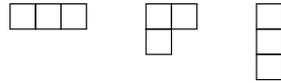
\begin{equation*}
    \ytableausetup{smalltableaux}
    \begin{ytableau}
        {} & {} & {}
    \end{ytableau}
    \qquad
    \ytableausetup{smalltableaux}
    \begin{ytableau}
        {} & {} \\
        {} 
    \end{ytableau}
    \qquad
    \ytableausetup{smalltableaux}
    \begin{ytableau}
        {} \\
        {} \\
        {}
    \end{ytableau}
\end{equation*}
of dimensions $1$, $2$, and $1$, respectively.
In the basis given by these irreps, $M$ takes the form
\[
    \begin{bmatrix}
    0 &             &           &           &           &\\
      & \frac{3}{2} &           &           &           &\\
      & &           \frac{3}{2} &           &           &\\
      & &           &           \frac{3}{2} &           &\\
      & &           &           &           \frac{3}{2} &\\
      & &           &           &           &           3\\
    \end{bmatrix}
\]

In general, one can show $\tfrac{d_{\lambda'_{*}}}{d_{\lambda}} \ge \tfrac{N - 2k}{N}$ which implies $e_{\lambda} \le 2k$ when the number of boxes of $\lambda$ below the first row is $k$.
Intuitively, this means that the subspace $\lambda \otimes \lambda'_{y}$ specifying that $y$ is not in the range occupies most of the dimension of $\lambda \otimes \lambda$, which is natural. This is essentially computed by the formula of $d_{\lambda}$ using the combinatorial property of Young diagram $\lambda$ (the \emph{hook length formula}, see \cref{FactFormulaHook}). Finally, with all things together, we have
$$\E[y]{ \tilde{p}_{k}^{y} }=\frac{1}{N}\norm{ \sqrt{M}\otimes I_{\bA} \ket{\psi_{k}} }^{2}\le \frac{1}{N}\max_{\operatorname{level}(\lambda) \le k} e_{\lambda}\le \frac{2k}{N},$$
which proves the average bound lemma as desired.

\paragraph{Summary of Our Technical Contribution} In summary, we highlight the main technical novelty of this work compared to previous papers.

Our main technical contribution is analyzing the permutation inversion problem in the bit-fixing model (see Lemma~\ref{LemBitfixing}) using representation theory inspired by Rosmanis~\cite{rosmanis:2022}. A key difficulty is that Rosmanis' techniques for permutation inversion without preprocessing do not directly apply to the permutation inversion problem in the bit-fixing model. In this model, we must separately bound (1) the advantage from queries made \emph{before} the challenge is revealed, and (2) the advantage from queries made \emph{after} the challenge is revealed. While Rosmanis' techniques extend naturally to handle the latter (online) queries, the main novelty of our proof lies in analyzing the former (offline) queries optimally by bounding the \emph{average information} that can be extracted by a bit-fixing algorithm in advance (see \cref{LemAvg}).

Concretely, since some points of the random permutation are effectively fixed before the challenge point is sampled, we must quantify how much information this can reveal about the challenge in advance. To do so, we introduce a specific linear operator $M$ acting on the oracle register to measure this information. We then apply representation theory of the symmetric group to decompose $M$ and to bound the maximal information gain about a random challenge point from fixing permutation points in advance.

Our key insight is to show that $M$ is compatible with the representation-theoretic structure of the symmetric group, which allows us to diagonalize it in a basis of irreducible representations (analogous to using a Fourier basis in standard matrix analysis). After diagonalization, we develop a novel method to bound the eigenvalues of $M$, in the spirit of the earlier analysis of Rosmanis using classical tools from representation theory such as Schur’s lemma and the branching rule. The spectral norm of $M$ then directly yields a bound on the advantage from the \emph{offline phase} (queries made before seeing the challenge). Combined with Rosmanis’ result on the \emph{online phase}, this gives our full upper bound for the bit-fixing model.

\subsection{Other Related Work}

In the quantum setting, apart from~\cite{nayebi2014quantum} and~\cite{hhan2019quantum}, other works have studied the permutation inversion problem in a different setting with 2-sided oracle access (namely with additional oracle access to $\pi^{-1}$ except at the challenge point). In the preprocessing setting, ~\cite{alagic2023two} obtained the same bound as \cite{nayebi2014quantum} and \cite{hhan2019quantum}. Without preprocessing, \cite{belovs2023one} showed separation between one-sided and 2-sided oracle access for the permutation inversion, and~\cite{cao2021being} gives several impossibility results for one-way permutations in the quantum setting, the most relevant to this work being the impossibility result that no quantum algorithm given 2-sided oracle access can invert a permutation with non-negligible probability by making less than $N^{1/5}$ queries.

In the classical setting, apart from Yao~\cite{Yao90}, ~\cite{Wee05, de2010time,coretti2018non} consider inverting $\epsilon$ fraction of input of the given permutation.  Wee~\cite{Wee05} proved the optimal lower bound of $ST=\tilde{\Omega}(N\epsilon)$ when $T=\tilde{O}(\sqrt{\epsilon N})$, and~\cite{de2010time,coretti2018non} proved the same bound for the full range of parameters.



\section{Proof of Main Theorem}\label{SecFramework}

The goal of this section is to prove \cref{ThmMain} below with the aid of the \emph{average bound} (\cref{LemAvg}), which will be proved in \cref{SecAvgbound} using the representation theory of symmetric groups.
In \cref{SecReduction},
    we rely on a modification of the main result in \cite{qipeng:2023} that reduces the success probability of an algorithm with auxiliary input \emph{i.e.} an algorithm with some preprocessed advice string (\cref{ThmMain}) to the success probability of a bit-fixing algorithm (\cref{LemBitfixing}).
In \cref{SecPurify}, we formally provide all necessary definitions related to bit-fixing algorithms.
In \cref{SecRecordNumber} and \ref{SecApprox},
    we briefly recall how \cite{rosmanis:2022} captures information from queries to a random permutation.
Finally, we will prove \cref{LemAvg} in \cref{SecBoundProof}.

Our main theorem below gives the tight upper bound of success probability of an auxiliary-input algorithm for permutation inversion:
\ThmMain*

\subsection{Reducing Game with Quantum Advice to Bit-Fixing Model}\label{SecReduction}

It is generally difficult to directly bound the success probability of algorithm generating some preprocessed advice.
Because of this, it is common to first show a reduction to another model without such advice,
    and then to prove the desired bound in the new model.
In \cite{kmsy:2020}, they propose a method of reducing auxiliary-input algorithms to multi-instance algorithms.
With this technique, they showed an $ST+T^2=\Omega(\epsilon N)$ bound for inverting $\epsilon$ fraction of inputs of a  random function with \emph{classical} advice. 
However, their approach suffers from a loss (of the exponent) on the security upper bound for \emph{quantum} advice, namely, they only showed a $ST+T^2=\Omega(\epsilon^3 N)$ bound.
This limitation arises from the unclonable nature of quantum advice.

In a later work, \cite{qipeng:2023}, Liu introduces a technique known as alternative measurements that allows us to use a single copy of advice when reducing from an Auxiliary-Input algorithm for function inversion to a bit-fixing algorithm for function inversion.
By using only a single copy of advice, this reduction yields $ST+T^2=\Omega(\epsilon N)$ bound for quantum advice.  

Our main goal is to prove the same security bound for the permutation inversion problem. We remark that although in the context of function inversion, it is a major open problem to close the gap between the above security bound and known algorithms (see~\cite{corrigan2019function}), the same security bound will be optimal for the permutation inversion problem (i.e., matching known attacks).


We extend his result to deal with random permutation below.
A complete proof is deferred to \cref{AppBitFix}.
Here we provide only a brief explanation of why the generalization holds.
Formally, we will reduce from auxiliary-input algorithms for permutation inversion to the following:
\begin{definition}[Bit-Fixing Model]\label{DefBitFixOWP}
A quantum algorithm $\cA$ for permutation inversion problem with $T$-queries in the \emph{$P$-bit-fixing model} ($(P,T)$- algorithm for short) consists of two stages:
\begin{itemize}
        \item[--] \textbf{Offline phase.} 
        \begin{enumerate}
            \item A uniformly random permutation $\pi \leftarrow S_N$ is sampled;
            \item The offline algorithm makes $P$ quantum queries to $\pi$, and outputs a bit $b$;
            \item If $b \neq 0$, the process restarts\footnote{We require that $\P{ b = 0 } > 0$.} from Step 1.
        \end{enumerate}
        \item[--] \textbf{Online phase.}
        \begin{enumerate}
            \item A uniformly random challenge $y \leftarrow [N]$ is sampled;
            \item Sharing the same inner register with the offline algorithm, the online algorithm takes $y$ as input, makes additional $T$ quantum queries to $\pi$ and outputs an answer $\ans$.
        \end{enumerate}
\end{itemize}
The algorithm succeeds if $\pi(\ans) = y$.
Denote by $\algo^{\pi}(y) \coloneq \ans$ the output of such an algorithm.
\end{definition}

Notice that the shared register can be understood as the online algorithm knows the strategy of the offline phase, which is formalized in \cite{qipeng:2023} as the offline algorithm outputs an additional quantum state $\tau$ and the online algorithm takes as an additional input.
We then rely on the following reduction, adapted from~\cite[Theorem~6.1]{qipeng:2023}:
\Reduction*

Given this reduction, our main theorem will follow directly from the following lemma, which we prove in \cref{SecBoundProof} below.

\Bitfixing*

\subsection{Formalism of Bit-Fixing Model}\label{SecPurify}

In this section we define the basic notation used in our proof. For convenience, we purify the bit-fixing model from two perspectives: the oracle register and the binary outcome of the offline phase. 

Instead of sampling the permutation $\pi$ in advance,
    we introduce an oracle register $\bO$ holding a state from the Hilbert space
\[ \cF = \Span \Big \{ \ket{\pi} \mid \pi \in \sym[N] \Big \} \]
spanned by all possible permutations of $N$ elements.
The register $\bO$ is initialized in the uniform superposition $\ket{v_{\emptyset}}$ over all permutations which is measured at the end of the computation to select a uniform permutation $\pi$.
A permutation inversion algorithm has an input register $\bX$ and an output register $\bY$ and makes a query by interacting with $\bO$ using the unitary
\[
    \orcl \coloneq \sum_{\substack{x, y \in [N] \\ \pi \in S_N}}
        \out{\pi}{\pi}_{\bO}
        \otimes \out{x}{x}_{\bX}
        \otimes \out{y \oplus \pi(x)}{y}_{\bY}.
\]

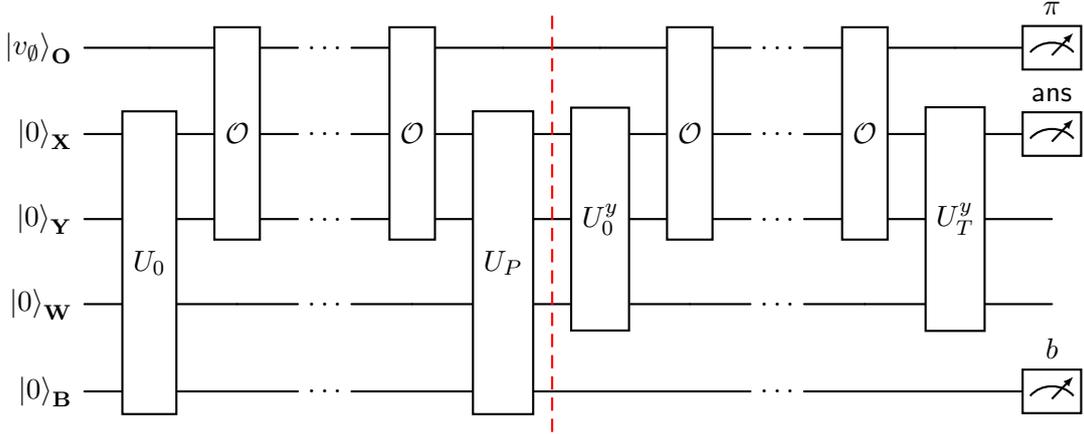
\begin{figure}[ht]
    \begin{center}
        \begin{quantikz}
            \lstick{$\ket{v_{\emptyset}}_\bO$} &&\gate[3]{\cO} & \ \ldots\ &\gate[3]{\cO}&&&\gate[3]{\cO}& \ \ldots\ &\gate[3]{\cO}&&\meter{\pi}\\
            \lstick{$\ket{0}_\bX$}&\gate[4]{U_0}& & \ \ldots\ &&\gate[4]{U_P}\slice{}&\gate[3]{U_0^y}&& \ \ldots\ &&\gate[3]{U_T^y}&\meter{\ans}\\
            \lstick{$\ket{0}_\bY$}&& & \ \ldots\ &&&&& \ \ldots\ &&&\\
            \lstick{$\ket{0}_\bW$}&& & \ \ldots\ &&&&& \ \ldots\ &&&\\
            \lstick{$\ket{0}_\bB$}&& & \ \ldots\ &&&&& \ \ldots\ &&&\meter{b}
        \end{quantikz}
    \end{center}
    \caption{The red dashed line indicates the point at which the algorithm $\cA$ receives the challenge $y$. The left part depicts the offline phase, and the right part depicts the online phase.}
    \label{FigBitFixing}
\end{figure}

We also introduce a working register $\bW$ for the algorithm and a single-qubit register $\bB$ which is measured (at the end of computation) to give the bit $b$ output by the offline stage. Let $\bA:=\bX\bY\bW \bB$ be the joint register. A $(P,T)$-alogrithm in \cref{DefBitFixOWP} can be formalized in the following way. See Figure \ref{FigBitFixing} for an overview illustration.

\paragraph{Offline phase.} The offline algorithm makes $P$ queries and computes a bit in the register $\bB$.
It is specified by unitaries $U_{0}, U_{1}, \dots, U_{P}$ acting on $\bA$. The computation starts in the state
\[
    \ket{\psi_{0}} := \ket{v_{\emptyset}}_{\bO} \ket{0}_{\bA}
\]
and the state at the end of the offline phase is
\[
    \ket{\psi_P} :=
         U_P
        \cO U_{P-1}
        \cO
        \cdots
        \cO
        U_0
        \ket{\psi_0}
\]
where unitaries $\cO$ and $U_i$ implicitly act on a larger Hilbert space than originally defined by tensoring with the identity operator on unaffected registers.
Measuring $\bB$ with outcome $b=0$ yields the postselected state $\ket{\phi_P}$.
Precisely, writing
\[
    \ket{\psi_P} = a_0\ket{z_0}\ket{0}_{\bB} + a_1\ket{z_1}\ket{1}_{\bB},
\]
we define $\ket{\phi_P} := \ket{z_0}\ket{0}_{\bB}$,
    which is well defined under the assumption that $\P{ b = 0 } > 0$.

\paragraph{Online phase.} After receiving a challenge $y \in [N]$, the online algorithm makes additional $T$ queries and computes an answer $\ans$ on the register $\bX$. It is specified by $y$-dependent unitaries $U^{y}_{0}, U^{y}_{1}, \dots, U^{y}_{T}$ on the joint register $\bX\bY\bW$ (the algorithm \emph{no longer} has the access to $\bB$).
The computation starts in $\ket{\phi_{P}}$. Let
\[
    \ket{\phi_{P+k}^{y}} \coloneq
        U^y_k \cO U^y_{k-1} \cO
        \cdots
        \cO U^y_0
        \ket{\phi_P}.
\]
for $k=0,1,\dots,T$ where $\cO$ and $U^y_i$ act similarly as above.

\paragraph{Success probability.}  
The success probability of such an algorithm is the expectation over $y \in [N]$ of the probability of measuring $\ket{\phi_{P + T}^{y}}$ yielding the outcome $\pi(\ans) = y$. A precise description of the projection of this measurement will be given in \cref{SecApprox}. 



\subsection{How to Record Number of Query}\label{SecRecordNumber}
Here, we define how the possible subspaces into which the permutation state $\ket{\pi}$ in the oracle register can change with each query made.

Given $k$ distinct points $x_{1}, \dots, x_{k} \in [N]$, we define a \emph{$k$-partial assignment} to be an injective function $\alpha : \{ x_{1}, \dots, x_{k} \} \rightarrow [N]$.
Then, define for any $k$-partial assignment $\alpha$ the subset
\[
    S_{\alpha} \coloneq \Big \{
        \pi \in \sym[N] \mid \pi(x_{i}) = \alpha(x_{i})~ \forall i \in [k]
    \Big \}
\]
of $S_N$ \emph{compatible} with $\alpha$.
Then, we define $\ket{v_{\alpha}}$ to be the uniform superposition over all permutations in $S_{\alpha}$. In particular, for $k = 0$, $\ket{v_\emptyset}$ is the uniform superposition over all $\pi \in \sym[N]$, as in \cref{SecPurify}.

Then, we may define the subspace of all $k$-partial assignments:
\[
    A_{k} \coloneq \Span \big \{
        \ket{v_{\alpha}} \mid |\alpha| = k
    \big \}.
\]
For convenience, for any $x, y \in [N]$ define the projection
\[ \Xi_{x}^{y} \coloneq \sum_{\pi \in \sym[N] :~\pi(x) = y} \out{\pi}{\pi}_{\bO} \]
on all permutations mapping $x$ to $y$. Then, the purified oracle is equivalently
\begin{equation}\label{EqOracle}
    \orcl \coloneq \sum_{x, y, z \in [N]} \Xi_{x}^{y}
        \otimes \out{x}{x}_{\bX}
        \otimes \out{z \oplus y}{z}_{\bY}.
\end{equation}

The following lemma, a reformulation of \cite[Theorem 3 (i)]{rosmanis:2022} and a permutation analogue of \cite[Lemma 2.3]{kmsy:2020},
    shows that after $k$ queries, the algorithm's state is entangled only with states in $A_{k}$.
\begin{lemma}\label{Lem1}
    For any algorithm with inner register $\bA$ having made $k$ queries to $\cO$, the joint state of the oracle and the algorithm is of form
    \[
        \sum_{v,z} \alpha_{v,z} \ket{v}_{\bO} \otimes \ket{z}_{\bA}
        \qquad\text{with}\qquad
        \sum_{v,z} | \alpha_{v, z} |^2 = 1
    \]
    where $\ket{v}$ is a chosen orthogonal basis\footnote{
        Unfortunately, there is no standard choice of an orthogonal basis of $A_{k}$.
        Although by definition $A_{k}$ is spanned by $\ket{v_{\alpha}}$ for all $k$-partial assignments $\alpha$,
            $\ket{v_{\alpha}}$ are neither a basis nor orthogonal to each other.
        } of $A_k$.
\end{lemma}
\proof By the proof of \cite[Theorem 3 (i)]{rosmanis:2022}, we can directly check from the definitions that $\Xi_x^y(A_k) \subset A_{k+1}$ for $k=0,\dots,N-1$ and any $x,y\in [N]$. The lemma is then immediate from the expression (\ref{EqOracle}) and the above with an inductive argument.\qed

\subsection{How to Record Success of Inverting a Challenge}\label{SecApprox}
To quantify success, define the \emph{success projection} for challenge $y$:
\[ P_{y} \coloneq \sum_{x \in [N]} \Xi_{x}^{y} \otimes \out{x}{x}_{\bX}\otimes I_{\bY\bW\bB}. \]
Then, using the notation of \cref{SecPurify}, a $(P,T)$-algorithm’s success probability on $y$ is defined as
\begin{equation}\label{EqWinNorm}
p^y_{\Succ} \coloneq \norm{ P_{y} \ket{\psi_{P + T}^{y, b = 0}} }^{2}
\end{equation}

To approximate $p^y_{\Succ}$, define for $k \in [N - 1]$ the subspace
\[
    A_{k}^{y} \coloneq \Span \Big \{
        \ket{v_{\alpha}} \mid |\alpha| = k \wedge y \in \im(\alpha)
    \Big \}
\]
and set $A_{0}^{y} \coloneq \{ 0 \}$ for any $y\in [N]$.
There is a nested chain
\[
    \Span \big \{ \ket{v_{\emptyset}} \big \} = A_{0}
        \subset A_{1}^{y}
        \subset A_{1}
        \subset \cdots
        \subset A_{N - 2}
        \subset A_{N - 1}^y
        = A_{N - 1}
        = \cF
\] (see \cite[Section 3.1]{rosmanis:2022} for details).
We define the \emph{high} and \emph{low} probability subspaces for inverting $y \in [N]$ as
\[
    \mathcal{H}^{\high}_{ y} = \bigoplus_{i = 1}^{N-1} \left (
        A_{i}^{y} \cap A_{i - 1}^{\perp}
    \right )
    \qquad\text{and}\qquad
    \mathcal{H}^{\low}_{y} = \bigoplus_{i = 0}^{N-1} \left (
        A_{i} \cap (A_{i}^{y})^{\perp}
    \right ),
\]
and define $\Pi^{\high}_y$ and $\Pi^{\low}_y$ their orthogonal projections, respectively. Intuitively, the high subspace consists of databases that obtain the full information of the preimage of $y$ exactly on the $i$-th query for some $i$. We refer \cite{rosmanis:2022} for more justification.

Recall $\bA := \bX \bY \bW \bB$ and define the projection $\widetilde{\Pi}^{\high}_{y} \coloneq \Pi^{\high}_{y} \otimes I_{\bA}$, which measures how much entanglement there is between the algorithm's state and information regarding the preimage of $y$ in the oracle register (in the subspace $\mathcal{H}_{y}^{\high}$). With the notation in \cref{SecPurify}, the following lemmas from \cite{rosmanis:2022} show that $\cH_y^{\high}$ approximates the true success subspace and that each additional query increases  amplitude by at most $\BO{ 1 / \sqrt{N} }$.
\begin{lemma}\cite[Theorem 3 (ii)]{rosmanis:2022}\label{Lem2}
    For each $y\in [N]$, we have
    \[
        \sqrt{p^{y}_{\Succ}} = \norm{
            P_{y} \ket{\phi^{y}_{P + T}}
        } \leq \norm{
            \widetilde{\Pi}^{\high}_{y} \ket{\phi^{y}_{P + T}}
        } + \frac{1}{\sqrt{N - 2(P + T)}}
    \]
\end{lemma}

\begin{lemma}\cite[Theorem 3 (iii)]{rosmanis:2022}\label{Lem3}
    For each $y\in [N]$ and $k \in [T]$, we have
    \[
        \norm{
            \widetilde{\Pi}_{y}^{\high} \ket{\phi^{y}_{P + k}}
        } \leq \norm{
            \widetilde{\Pi}_{y}^{\high} \ket{\phi^{y}_{P + k - 1}}
        } + \frac{2 \sqrt{2}}{\sqrt{N - 4(P + k)}}
    \]
\end{lemma}

\begin{remark}
Our definition of the high subspace differs slightly from that in~\cite{rosmanis:2022}, where the author sets
\[
    \cH^{\high}_k := \bigoplus_{i = 1}^{k} \bigl( A_i^{0} \cap A_{i-1}^{\perp} \bigr),
\]
with the challenge fixed to $0$ and uses the notation $B_k := A_k^0$. In particular, $\cH^{\high}_k$ there depends on the query number $k$. To avoid confusion with indices, we instead denote this subspace by $\cH^{\high}_{0,k}$, and write $\Pi^{\high}_{0,k}$ for its orthogonal projection and $\widetilde{\Pi}^{\high}_{0,k} := \Pi^{\high}_{0,k} \otimes I_{\bA}$.

This distinction does not affect Lemma~\ref{Lem2} and Lemma~\ref{Lem3}.
First, the results extend to any fixed challenge $y \in [N]$, not just $0$.
Moreover, since any $\ket{v} \in A_k$ is orthogonal to $A_i^0 \cap A_{i-1}^{\perp}$ for all $i>k$, we have $
    \norm{ \widetilde{\Pi}^{\high}_{0} \ket{\phi^{0}_{P + T}} }
    = \norm{ \widetilde{\Pi}^{\high}_{0, P + k} \ket{\phi^{0}_{P + k}} }
$ for all $k \in [T]$, as $\ket{\phi^{0}_{P + k}} \in A_{P + k}$.
Hence, all relevant norms coincide with those in~\cite[Theorem~3]{rosmanis:2022}.
\end{remark}

\subsection{Upper Bound of Permutation Inversion in Bit-Fixing Model}\label{SecBoundProof}

With all necessary notation defined, we are now ready to prove \cref{LemBitfixing}.
In this proof, we will use our main bound on the average information captured by the projection $\Pih$ defined in the previous section, which we state here:
\begin{restatable}[Average Bound]{lemma}{Avg}\label{LemAvg}
    For any $\ket{v} \in A_k$, we have
    \[ \E[y][\bigg]{ \norm{ \phigh_{y} \ket{v} }^{2} } \leq \frac{2k}{N} \]
    where the expectation is over a uniformly random challenge $y \in [N]$.
\end{restatable}

We will prove this lemma using the \emph{representation theory} of symmetric groups in \cref{SecAvgbound}.
Intuitively, the expectation over all challenges exhibits invariance properties that can be captured and exploited via representation theory. Assuming this lemma holds, we prove the following bound:
\Bitfixing*

\proof As previous sections, we can model the success probability of a $(P,T)$-algorithm $\cA$ as the expectation of a measurement on the overall state, postselected on the outcome $b = 0$ (see \eqref{EqWinNorm}):
\begin{equation}\label{EqWinNorm2}
    \P[\pi, y][\Big]{
        \pi \left ( \algo^{\pi}(y) \right ) = y
    } = \E[y]{ p^{y}_{\Succ} } = \E[y][\bigg]{
        \norm{ P_y\ket{\phi_{P+T}^{y}} }^2
    }.
\end{equation}

For any challenge $y \in [N]$, the quantity $\norm{ P_{y} \ket{\phi_{P + T}^{y}} }$ can be approximated with \cref{Lem2} by the projection $\Pi_y^{\high}$ defined in \cref{SecApprox}.
Together with using \cref{Lem3} recursively, we obtain 
\begin{align*}
    \norm{ P_{y} \ket{\phi_{P + T}^{y}} } \le \norm{
        \widetilde{\Pi}_y^{\high} \ket{\phi_{P}^{y}}
    } + \frac{2 \sqrt{2}(T + 1)}{\sqrt{N - 4(P + T)}}.
\end{align*}
Notice that $\ket{\phi_{P}^{y}}$ is obtained from $\ket{\phi_P}$ by applying the unitary $U_{0}^{y}$, and $\widetilde{\Pi}_y^{\high} = \Pi_y^{\high} \otimes I_{\bA}$ commutes with $U_0^y$. Hence,
\[
    \norm{
        \widetilde{\Pi}_{y}^{\high} \ket{\phi_{P}^{y}}
    } = \norm{
        \widetilde{\Pi}_y^{\high} \ket{\phi_{P}}
    }
\]
where we replace $\ket{\phi_{P}^{y}}$ in the norm by $y$-independent $\ket{\phi_{P}}$.
Applying the arithmetic-geometric inequality, this gives
\[
    \norm {
        P_y\ket{\phi_{P+T}^{y}}
    }^2 \le 2 \norm{
        \widetilde{\Pi}_y^{\high} \ket{\phi_{P}}
    }^2 + \frac{16(T + 1)^2}{N - 4(P + T)}.
\]
And substituting this into \eqref{EqWinNorm2}, we have
\[
    \P[\pi, y]{
        \pi \left ( \algo^{\pi}(y) \right ) = y
    } \le 2\E[y][\bigg]{
        \norm{
            \widetilde{\Pi}_{y}^{\high} \ket{\phi_{P}}
        }^2
    } + \frac{16(T + 1)^2}{N - 4(P + T)}.
\]

By definition, $\ket{\phi_P}$ is obtained by measuring $\ket{\psi_P}$ on the register $\bB$ and postselecting on the outcome $b = 0$.
From \cref{Lem1}, the state $\ket{\psi_{P}}$ after $P$ offline queries, and hence $\ket{\phi_P}$, can be written as
\[
    \sum_{v,z} \alpha_{v,z} \ket{v}_{\bO} \otimes \ket{z}_{\bA}
    \qquad\text{with}\qquad
    \sum_{v,z} | \alpha_{v,z} |^2 = 1
\]
where and all $\ket{v} \in A_P$ and we assume that the above coefficients hold for $\ket{\phi_P}$.
Therefore,
\begin{align*}
    \E[y][\bigg]{
        \norm{
            \widetilde{\Pi}_y^{\high}\ket{\phi_{P}}
        }^2
    } &= \frac{1}{N} \sum_{y\in [N]} \norm{
        \widetilde{\Pi}_y^{\high} \ket{\phi_{P}}
    }^2\\&= \frac{1}{N} \sum_{y \in [N]} \norm{
        \sum_{v, z} \alpha_{v, z} \Pi^{\high}_y \ket{v}_{\bO} \otimes\ket{z}_{\bA}
    }^2\\&= \frac{1}{N} \sum_{v,z} | \alpha_{v,z} |^2
        \sum_{y \in [N]} \norm{\Pi^{\high}_{y} \ket{v}_{\bO} \otimes\ket{z}_{\bA}}^2\\
    &= \frac{1}{N} \sum_{v,z} | \alpha_{v,z} |^2
        \sum_{y \in [N]} \norm{
            \Pi^{\high}_y\ket{v}
        }^2\\
    &= \sum_{v,z} | \alpha_{v,z} |^2 \E[y][\bigg]{
        \norm{
            \Pi_{y}^{\high} \ket{v}
        }^2
    } \le \frac{2P}{N}
\end{align*}
where the last inequality uses Lemma \ref{LemAvg} and $\sum_{v,z} | \alpha_{v,z} |^2 = 1$.

Consequently,
\[
    \P[\pi, y][\Big]{
        \pi \left ( \algo^{\pi}(y) \right ) = y
    }\le \frac{4P}{N} + \frac{16(T + 1)^2}{N - 4(P + T)} = \BO{
        \frac{P}{N} + \frac{T^{2}}{N}
    }
\]
for $P + T \ll N$.
Otherwise, if $P + T = \Omega(N)$, the bound becomes trivial.\qed

\section{Proof of Average Bound}\label{SecAvgbound}
In this section, we prove the average bound lemma used previously. The proof relies on the representation theory of the symmetric group. A brief review of the necessary background is provided in \cref{AppRep}. Readers already familiar with representation theory may skip directly to the proof and refer back to the appendix as needed. The average bound is stated as:

\Avg*

In order to bound the expectation, we define the following operator
\[ M \coloneq \sum_{y \in [N]} \Pih  \]
which is the sum of all projections on the subspace of inverting $y$ with high probability.
Note that for each $y \in [N]$, $\Pih$ is an orthogonal projection,
    and thus positive semi-definite.
As such, while $M$ defined above is \emph{not} itself a projection,
    it is still positive semi-definite,
    and thus is both self-adjoint and has a unique matrix square root.
Having defined $M$ as above, we observe that
\begin{align*}
    N\cdot\E[y]{ \| \Pih \ket{v} \|^{2} } &=
    \sum_{y \in [N]} \| \Pih \ket{v} \|^{2}= \sum_{y \in [N]} \bra{v} \Pih \ket{v}\\
        &= \bra{v} \sum_{y \in [N]} \Pih \ket{v}
         = \bra{v} \sqrt{M}^{\dagger} \sqrt{M} \ket{v}
         = \| \sqrt{M} \ket{v} \|^{2},
\end{align*}
merging all $\Pih$ together into $M$,
and hence the lemma follows by showing for all $\ket{v} \in A_{k}$ that
\begin{equation}\label{EqMbound}
    \norm{\sqrt{M} \ket{v}}^2 \le 2k
\end{equation}

We organize the rest of the proof as follows.
In \cref{SecDecomp}, we recall the notation of representation theory necessary to understand our proof, as well as the specific results from \cite{rosmanis:2022} about decomposing various subspaces we are interested in.
In \cref{SecSym}, we then show that $M$ maps between representations of symmetric groups in a ``compatible'' way, and thus can be written as a block diagonal matrix in a particular basis.
Finally, we use multiple equivalent decompositions of $M$ to compute its eigenvalues in \cref{SecMainProof}, concluding the proof of the average bound.

\subsection{Decomposition of Database Space via Representation}\label{SecDecomp}

We have already defined the subspace $A_{k}$ of databases after $k$-queries and the high subspace $\cH^{\high}_{y}$ of inverting $y$.
\cite{rosmanis:2022} observes that they can be decomposed into \emph{irreducible representation of symmetric groups}.
We recall necessary notation and results in this section and will see later why such decompositions help us in \cref{SecSym}.

Briefly speaking, a group representation (\cref{DefRep}) is a Hilbert space with an additional structure of a group action.
An irreducible representation (\cref{DefIrrep}, ``irrep'' for short) is a minimal representation that cannot be further decomposed into subrepresentations (\cref{DefSubrep}, namely subspaces preserved by the action), and every representation can be decomposed into irreps up to isomorphism; they play the central role for our proof.
We mostly focus on representations of symmetric groups or products of symmetric groups.
In this section, we use the notation $S_{[N]}$ for $S_N$ and consider the subgroups
\[
    S_{[N] \setminus \{y\}} \coloneq \left \{
        \pi \in S_{[N]} \mid \pi(y) = y
    \right \}
\]
of permutations fixing $y$ for each $y\in [N]$.
Irreducible representations of $S_{[N]}$ are in one-to-one correspondence with Young diagrams of size $N$ (\cref{DefYoung}) up to isomorphism (\cref{FactCorres}).
This fact plays a central role in how we decompose database-like states in $A_{k}$.
By abuse of notation, we will interchangeably refer to an irrep (and its underlying Hilbert space) by the corresponding Young diagram.

We additionally introduce specialized notation for dealing with Young diagrams.
Fix $0 \le k < N$ and $y\in [N]$.
Let $\theta$ be a Young diagram of size $k$, denoted $\theta \vdash k$.
Define
\[ \overline{\theta} \coloneq (N - k, \theta) \]
to be a Young diagram of size $N$ (constructed by adding a row of length $(N - k)$ above $\theta$), and define
\[ \overline{\theta}_{*} \coloneq (N - k - 1, \theta) \]
to be a Young diagram of size $N - 1$ constructed similarly.
To ensure that $\overline{\theta}$ and $\overline{\theta}_{*}$ are well defined Young diagrams,
    we always assume that $k \le \frac{N}{2}$.
If this is not the case, the permutation inversion problem becomes trivial,
    so this assumption is made without loss of generality for our proof.
As noted above, $\overline{\theta}$ is also used to denote the corresponding irrep of $S_{[N]}$,
    and $\overline{\theta}_y$ is used to denote the irrep of $S_{[N] \setminus \{y\}}$
    (corresponding to $\overline{\theta}_{*}$).
Here, the lower index of $\overline{\theta}_{y}$ is used to specify the point removed from $[N]$.

\begin{figure}[ht]
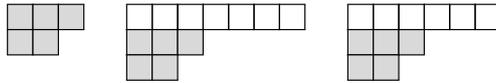

    \centering
    \begin{tabular}{ c c c }
        \ytableausetup{smalltableaux}
        \begin{ytableau}
            *(gray!30) & *(gray!30) & *(gray!30) \\
            *(gray!30) & *(gray!30)
        \end{ytableau} & 
        \ytableausetup{smalltableaux}
        \begin{ytableau}
            {} & {} & {} & {} & {} & {} & {} \\
            *(gray!30) & *(gray!30) & *(gray!30) \\
            *(gray!30) & *(gray!30)
        \end{ytableau} &
        \ytableausetup{smalltableaux}
        \begin{ytableau}
            {} & {} & {} & {} & {} & {} \\
            *(gray!30) & *(gray!30) & *(gray!30) \\
            *(gray!30) & *(gray!30)
        \end{ytableau} 
    \end{tabular}
    \caption{
        Example of $\theta$, $\overline{\theta}$, and $\overline{\theta}_{*}$ (left to right) for $N = 12$, $k = 5$.
    }
    \label{FigBar}
\end{figure}

The Hilbert space on the oracle register ($\cF$ above) can be regarded as the regular representation (\cref{AppSecReg}) of $S_{[N]} \times S_{[N]}$ equipped with an additional structure of group action
\begin{align}\label{EqRegular}
    V : S_{[N]} \times S_{[N]} &\rightarrow \U(\cF)\\
        (\pi_{D}, \pi_{R}) &\mapsto V_{\pi_{D}}^{\pi_{R}} : \cF \rightarrow \cF \notag
\end{align}
where $V_{\pi_{D}}^{\pi_{R}}$ is a unitary defined on the computational basis mapping $ \ket{\pi} \mapsto \ket{\pi_{R} \circ \pi \circ \pi_{D}^{-1}}$ for any $\pi \in S_{[N]}$.

For any irrep $\lambda$ of $S_{[N]}$, we denote $\lambda \otimes \lambda$ the tensor product of two copies of $\lambda$, which is an irrep of $S_{[N]} \times S_{[N]}$ (\cref{FactProduct}).
It follows from the properties of the regular representation (\cref{FactRegIsoty}) that we can decompose $\cF$ as
\begin{equation}\label{EqRegIsoty}
    \cF = \bigoplus_{\lambda \vdash N} \cH_{\lambda}
\end{equation}
where the sum is over all Young diagrams of size $N$ and $\cH_{\lambda}$ is a subrepresentation of $\cF$ isomorphic to $\lambda \otimes \lambda$.

Importantly, notice for any $k$-partial assignment $\alpha$, applying $V_{\pi_{D}}^{\pi_{R}}$ to $\ket{v_{\alpha}}$ simply permutes the domain and range of $\alpha$, yielding $\ket{v_{\beta}} \in A_{k}$ for $\beta \coloneq \pi_{R} \circ \alpha \circ \pi_{D}^{-1}$ (which is also a $k$-partial assignment).
Thus, $A_{k}$ can be decomposed as a subset of the decomposition in \cref{EqRegIsoty}.
Specifically, the following holds:

\begin{lemma}\cite[Claim 13]{rosmanis:2014}\label{PropDecompA}
    For any $k \geq 0$, there is an orthogonal decomposition
    $$A_k=\bigoplus_{\theta\in Y_{\le k}}\cH_{\overline{\theta}}$$
    where $Y_{\le k}$ is the set of Young diagrams of size $\le k$ such that $\overline{\theta}$ is well defined.
\end{lemma}

Similarly, for each $y$, the subspaces $\cH^{\high}_{y}$ and $\cH^{\low}_{y}$ are representations, and can therefore be decomposed into irreps.
However, in order to find a decomposition that will be useful in our proof below,
    we first need a finer analysis of (\cref{EqRegIsoty}).
In this pursuit, we examine the representation of
    $S_{[N]} \times S_{[N] \setminus \{y\}}$
    defined on $\cF$ via the natural inclusion
    $S_{[N]} \times S_{[N] \setminus \{y\}} \hookrightarrow S_{[N]} \times S_{[N]}$,
    which in general is called the restricted representation (\cref{DefRestrict}).
Intuitively, this subgroup acting on the same space is in some sense a ``coarser'' action,
    and, as such, when decomposing into irreps, this coarser structure results
    in a ``finer'' decomposition into irreps.

Formally, the Braching rule (\cref{FactBranch}) asserts that for any Young diagram $\lambda \vdash N$,
    restricting the corresponding $S_{[N]}$-irrep $\theta$ to a
    $S_{[N] \setminus \{y\}}$-representation allows $\lambda$ to be further decomposed into 
    $S_{[N] \setminus \{y\}}$-irreps corresponding to Young diagrams obtained from $\lambda$ by removing one box (denoted $\mu \prec \lambda$).
Then, by \cref{FactProduct}, $\lambda \otimes \mu$ is an irrep of
    $S_{[N]} \times S_{[N]\setminus\{y\}}$.
Specifically, we have a decomposition
\begin{equation}\label{EqBranchLambda}
    \lambda = \bigoplus_{\mu \prec \lambda} \mu
\end{equation}

And in particular, when $\lambda$ is of form $\overline{\theta}$ for some $\theta \vdash k$,
    by considering the restriction of $\lambda$ as a $S_{[N]} \times S_{[N]\setminus\{y\}}$-representation,
    we find that
\begin{equation}\label{EqBranchIsoty}
    \cH_{\overline{\theta}} = \left(
        \bigoplus_{\rho \prec \theta} \cH_{\overline{\theta}}^{\overline{\rho}_y}
    \right ) \oplus \cH_{\overline{\theta}}^{\overline{\theta}_y}
\end{equation}
where $\cH_{\overline{\theta}}^{\overline{\rho}_y}$ and
    $\cH_{\overline{\theta}}^{\overline{\theta}_y}$ are subrepresentations of
    $\cH_{\overline{\theta}}$ isomorphic to $\overline{\theta} \otimes \overline{\rho}_y$ and
    $\overline{\theta}\otimes\overline{\theta}_y$, respectively.

As mentioned above, $A_{k}$ is a $S_{[N]} \times S_{[N]}$ representation.
For any $k$-partial assignment $\alpha$ such that $y \in \operatorname{Im}(\alpha)$,
    it holds that
    $V_{\pi_{D}}^{\pi_{R}}\ket{v_{\alpha}} = \ket{v_{\beta}}$
    for $\beta \coloneq \pi_{R} \circ \alpha \circ \pi_{D}^{-1}$.
Importantly here, notice that $y$ is still in the image of $\beta$.
This fact directly implies that $A^{y}_{k}$ is a $S_{[N]} \times S_{[N] \setminus \{y\}}$ subrepresentation \footnote{
    However, it's \emph{not} a $S_{[N]} \times S_{[N]}$-subrepresentation of $\cF$ because arbitrary $g : S_{[N]} \rightarrow S_{[N]}$ cannot guarantee $y \in \im(\beta)$.
}  of $A_{k}$.

Since $\cH^{\high}_y$ and $\cH^{\low}_y$ are direct sums of various $A_{k}$ and $A_{k}^{y}$,
    then, they will naturally decompose as direct sums of the same form as \cref{EqBranchIsoty}.
The exact form of this decomposition is as follows:
\begin{lemma}\cite[Corollary 14]{rosmanis:2014}\label{PropDecompH}
    For any $k \geq 0$,
    \[
        \cH^{\high}_{y} \coloneq \bigoplus_{\theta\in Y_{\le N}}
        \bigoplus_{\rho \prec \theta} \cH_{\overline{\theta}}^{\overline{\rho}_y}
        \qquad\text{and}\qquad
        \cH^{\low}_y = \bigoplus_{\theta \in Y_{\le N}}
        \cH_{\overline{\theta}}^{\overline{\theta}_y}
    \]
\end{lemma}

For any Young diagram $\theta \vdash k$ and $\rho \prec \theta$, we denote 
    $\Pi_{\overline{\theta}}$, $\Pi_{\overline{\theta}}^{\overline{\rho}_{y}}$ and
    $\Pi_{\overline{\theta}}^{\overline{\theta}_{y}}$ the orthogonal projections on the subspaces $\cH_{\overline{\theta}}$,
    $\cH_{\overline{\theta}}^{\overline{\rho}_y}$ and
    $\cH_{\overline{\theta}}^{\overline{\theta}_y}$, respectively.
By orthogonality of those subspaces, if $\theta \neq \theta'$,
\begin{equation}\label{EqOrthProj}
    \Pi_{\overline{\theta}} \Pi_{\overline{\theta}}^{\overline{\rho}_{y}} =
        \Pi_{\overline{\theta}}^{\overline{\rho}_y}
    \qquad\text{and}\qquad
    \Pi_{\overline{\theta}} \Pi_{\overline{\theta'}}^{\overline{\rho'}_{y}} = 0
\end{equation}

\subsection{How to Use Symmetry of Average}\label{SecSym}

In this section, we explain how to leverage the symmetry of the operator $M$,
    which is defined as a sum over $y \in [N]$.
Because of this summation, $M$ possesses a natural invariance under permutations,
    which allows us to more easily analyze it through representation theory.
In particular, $M$ is a homomorphism of $S_{[N]} \times S_{[N]}$-representations (\cref{CorHomom}),
    and by applying the decomposition results from the previous section,
    we are able to diagonalize $M$ in a very convenient way.

Recall the regular representation $V^{\pi_{R}}_{\pi_{D}}$ defined in (\cref{EqRegular}).
The following lemma records an immediate consequence of the definition:
\begin{lemma}[Change of Challenge]\label{ChangeCoord}
    Let $y \in [N]$ and $\pi_{D},\pi_{R} \in S_{[N]}$.
    Then
    \[
        V^{\pi_{R}}_{\pi_{D}} \, \Pi^{\high}_y \, (V^{\pi_{R}}_{\pi_{D}})^{-1}
            = \Pi^{\high}_{\pi_{R}(y)}
    \]
\end{lemma}

\begin{proof}
Recall that $\cH^{\high}_{y}$ is defined as
\[
    \mathcal{H}^{\high}_{y}
    \coloneq \left ( A_{1}^{y} \cap A_{0}^{\perp} \right ) \oplus \cdots
        \oplus \left ( A_{N-1}^{y} \cap A_{N-2}^{\perp} \right )
\]
Thus it suffices to show that $V^{\pi_{R}}_{\pi_{D}}$ is a bijection from each component 
    $A_{i}^{y} \cap A_{i-1}^\perp$ to
    $A_i^{\pi_{R}(y)} \cap A_{i - 1}^\perp$ for every $i \in [N - 1]$.

For every basis vector $\ket{v_{\alpha}} \in A_{k}$,
    define $\beta \coloneq \pi_{R} \circ \alpha \circ \pi_{D}^{-1}$.
Then, $V^{\pi_{R}}_{\pi_{D}} \ket{v_{\alpha}} = \ket{v_{\beta}}$,
    and if $y \in \im(\alpha)$, then $\pi_{R}(y) \in \im(\beta)$.
Hence, $V^{\pi_{R}}_{\pi_{D}}$ is an bijection from $A_{i}^{y}$ to $A_{i}^{\pi_{R}(y)}$. 

Moreover, since $A_{i - 1}$ is itself a subrepresentation of $\cF$, 
$A_{i - 1}$ is be preserved by $V^{\pi_{R}}_{\pi_{D}}$,
    and since $V^{\pi_{R}}_{\pi_{D}}$ is a unitary representation,
    the same is true for the orthogonal complement.

Therefore, as desired,
\[
    V^{\pi_{R}}_{\pi_{D}} \big ( A_i^y \cap A_{i - 1}^{\perp} \big ) =
        A_{i}^{\pi_{R}(y)} \cap A_{i - 1}^{\perp}
\]

As a result, $V^{\pi_{R}}_{\pi_{D}}$ is a bijection from $\cH^{\high}_y$ to $\cH^{\high}_{\pi_{R}(y)}$,
    so if we define
\[
    \Pi \coloneq V^{\pi_{R}}_{\pi_{D}} \Pi^{\high}_y (V^{\pi_{R}}_{\pi_{D}})^{-1},
\]
then $\Pi$ is an orthogonal projection (clearly $\Pi^{2} = \Pi$) with image $\cH^{\high}_{\pi_{R}(y)}$,
    so we may conclude that $\Pi = \Pi^{\high}_{\pi_{R}(y)}$ as desired.
\end{proof}

As an immediate result, we obtain the key symmetry property of $M$:

\begin{corollary}\label{CorHomom}
The operator $M : \cF \to \cF$ is a homomorphism of $S_{[N]} \times S_{[N]}$-representations (\cref{DefGlinear}).
That is, for any $\pi_{D}, \pi_{R} \in S_{[N]}$,
\[
    M \circ V^{\pi_{R}}_{\pi_{D}} = V^{\pi_{R}}_{\pi_{D}} \circ M
\]
\end{corollary}

\begin{proof}
By definition of $M$,
\[
    V^{\pi_{R}}_{\pi_{D}} \circ M \circ (V^{\pi_{R}}_{\pi_{D}})^{-1} =
        V^{\pi_{R}}_{\pi_{D}}
            \circ \left ( \sum_{y \in [N]} \Pi^{\high}_y \right )
            \circ (V^{\pi_{R}}_{\pi_{D}})^{-1}
    = \sum_{y \in [N]} \Pi^{\high}_{\pi_{R}(y)} = M,
\]
where the second equality uses Lemma \ref{ChangeCoord} and the linearity of each $\Pi_{y}^{\high}$. This proves the claim.
\end{proof}

We can now use the fact that $M$ is a homomorphism of $S_{[N]} \times S_{[N]}$ representations
    to apply various results from the representation theory of symmetric groups.

To briefly recall, \cref{EqRegIsoty} views $\cF$ as the regular representation of $S_{[N]} \times S_{[N]}$ and shows the decomposition
\[ \cF = \bigoplus_{\lambda \vdash N} \cH_{\lambda} \]

By finding a basis of $\cF$ compatible with this decomposition, we can use Schur's lemma (
    see \cref{FactSchur} and \cref{AppSecReg} for a more detailed explanation
) to write $M$ as a diagonal matrix with blocks of eigenvalues corresponding to each irrep of $S_{[N]} \times S_{[N]}$.
Precisely, after changing basis, $M$ takes the form
\begin{equation}\label{EqBlockDiagonal}
    M = \sum_{\lambda \vdash N} e_{\lambda} \Pi_{\lambda},
\end{equation}
where each $e_{\lambda}$ is the only eigenvalue of $M$ on the subspace $\cH_{\lambda}$ with multiplicity equal to the dimension of the corresponding irrep $\lambda$.

\subsection{Proof Completion}\label{SecMainProof}

In this section we complete the proof of the average bound (\cref{LemAvg}) by proving \cref{EqMbound}. Namely, we show for any $\ket{v} \in A_{k}$ that
\[ \norm{\sqrt{M} \ket{v}}^2 \le 2k \]

From \cref{EqBlockDiagonal} in the previous section, we know that $M$ can be written in block-diagonal form with eigenvalues $e_{\lambda}$ corresponding to each irrep $\cH_{\lambda}$.
Moreover, from \cref{PropDecompA}, the space $A_{k}$ decomposes into subspaces $\cH_{\overline{\theta}}$ indexed by Young diagrams $\theta$ of size $\le k$.
We show uniformly over $\theta \vdash k$ that $e_{\overline{\theta}} \le 2k$.

Given this bound, for any $\ket{v} \in A_{k}$, we have the decomposition $\ket{v} = \sum_{\theta \in Y_{\le k}} a_{\theta} \ket{v_\theta}$ where $\ket{v_\theta} \in \cH_{\overline{\theta}}$ and $\sum_{\theta\in Y_{\le k}} | a_{\theta} |^2 = 1$,
    from which it holds that
\[
    \norm{ \sqrt{M} \ket{v} }^2
        = \sum_{\theta \in Y_{\le k}} |a_\theta|^2 e_{\overline{\theta}}
        \le 2k
\]

\paragraph{Step 1: Two Expressions for $M$.}  
To evaluate $e_{\overline{\theta}}$, we compare two ways of writing $M$.  
From (\ref{EqBlockDiagonal}), we have $M=\sum_{\lambda\vdash N} e_{\lambda} \, \Pi_{\lambda}$. On the other hand, by expanding each $\Pi_y^{\high}$ via Lemma \ref{PropDecompH},
\[
M=\sum_{y\in [N]} \Pi^{\high}_y
  =\sum_{y\in [N]}\sum_{\theta\in Y_N}\sum_{\rho\prec\theta} \Pi_{\overline{\theta}}^{\overline{\rho}_y}.
\]
Pre-composing both sides with $\Pi_{\overline{\theta}}$, all terms with $\theta'\neq\theta$ vanish by (\ref{EqOrthProj}).  
Thus we obtain
\begin{equation}\label{EqMTwoForm}
    e_{\overline{\theta}}\Pi_{\overline{\theta}}
    = \sum_{y\in [N]} \sum_{\rho\prec\theta} \Pi_{\overline{\theta}}^{\overline{\rho}_y}.
\end{equation}

\paragraph{Step 2: Dimension Counting.}  
We now evaluate $e_{\overline{\theta}}$ by comparing dimensions.  
Let $d_{\overline{\theta}}$ denote the dimension of the irrep $\overline{\theta}$ of $S_{[N]}$, and similarly let $d_{\overline{\theta}_*}$ and $d_{\overline{\rho}_*}$ denote the dimensions of the irreps $\overline{\theta}_y$ and $\overline{\rho}_y$ of $S_{[N]\setminus \{y\}}$.  
By the branching rule (Fact \ref{FactBranch}),
\[
\overline{\theta} \;\cong\; \Big(\bigoplus_{\rho\prec\theta} \overline{\rho}_y\Big)\;\oplus\;\overline{\theta}_y 
\qquad\text{and}\qquad
d_{\overline{\theta}} = \sum_{\rho\prec\theta} d_{\overline{\rho}_*} + d_{\overline{\theta}_*}.
\]

Furthermore, $\operatorname{rank}(\Pi_{\overline{\theta}})=d_{\overline{\theta}}^2$ and $\operatorname{rank}(\Pi_{\overline{\theta}}^{\overline{\rho}_y})=d_{\overline{\theta}}\, d_{\overline{\rho}_*}$ since $\cH_{\overline{\theta}}\cong \overline{\theta}\otimes \overline{\theta}$ and $\cH_{\overline{\theta}}^{\overline{\rho}_y}\cong \overline{\theta}\otimes \overline{\rho}_y$. Taking traces in (\ref{EqMTwoForm}) yields
\[
e_{\overline{\theta}} d_{\overline{\theta}}^2
  = \sum_{y\in [N]} \sum_{\rho\prec\theta} d_{\overline{\theta}}\, d_{\overline{\rho}_*}
  = N d_{\overline{\theta}} \sum_{\rho\prec\theta} d_{\overline{\rho}_*}
  = N d_{\overline{\theta}} \big(d_{\overline{\theta}}-d_{\overline{\theta}_*}\big).
\]
Dividing through by $d_{\overline{\theta}}^2$, we obtain
\[
e_{\overline{\theta}}
   = N\left(1-\frac{d_{\overline{\theta}_*}}{d_{\overline{\theta}}}\right).
\]

\paragraph{Step 3: Bounding $e_{\overline{\theta}}$.}
To properly bound $e_{\overline{\theta}}$, it remains to bound the ratio $d_{\overline{\theta}_*}/d_{\overline{\theta}}$.  
The following lemma provides the necessary estimate:

\begin{restatable}{lemma}{IneqDim}\label{IneqDim}
For any $\theta\vdash k$, we have $\tfrac{d_{\overline{\theta}_*}}{d_{\overline{\theta}}} \;\ge\; \tfrac{N-2k}{N}$.
\end{restatable}

Combining this with the previous expression gives
\[
e_{\overline{\theta}}
   = N\Big(1-\frac{d_{\overline{\theta}_*}}{d_{\overline{\theta}}}\Big)
   \le 2k,
\]
as required.  

The inequality of \cref{IneqDim} was first proved in \cite[Claim 7]{rosmanis:2014}.
For completeness, we provide a proof in \cref{AppRep} using the \emph{hook length formula} (\cref{FactFormulaHook}), which computes dimensions of irreps from the combinatorics of Young diagrams.

\ifec
    \bibliographystyle{splncs04}
\else
    \bibliographystyle{alpha}
\fi

\bibliography{refs}

\appendix

\section{Auxiliary Algorithm vs.\ Bit-Fixing Model}\label{AppBitFix}

Following the same strategy of \cite{qipeng:2023} for the quantum random function oracle (QROM), we reduce the security against adversary with auxiliary input to the security against adversary in the \emph{bit-fixing model}, now in the quantum random permutation model.

The previous approach of \cite{kmsy:2020} proceeds via the \emph{multi-instance game}, showing reductions from auxiliary algorithm to multi-instance game, and then to the computation of some form of conditional probability (captured by general bit-fixing model defined in later works). This yields tight time-space tradeoffs with \emph{classical} advice.
However, the argument does not extend tightly to \emph{quantum} advice because the advice cannot be cloned. To address this, \cite{qipeng:2023} introduced the \emph{alternating measurement game}, which preserves the single copy of advice and achieves tight bounds with quantum advice.

Our contribution here is a direct adaptation of the main result of \cite[Theorem~6.1]{qipeng:2023} and the observation that the proof is agnostic to the oracle distribution: the arguments apply verbatim when the oracle is a random permutation rather than a random function. We focus only on the permutation inversion problem and refer the general reduction to \cite{qipeng:2023}.

Let $\delta(S,T)$ and $\nu(P,T)$ be the maximum success probability of all adversary with $S$-qubit advice and $T$-quantum queries (in Theorem \ref{ThmMain}) and all adversary with $T$-quantum queries in the $P$-bit-fixing model (in Definition \ref{DefBitFixOWP}), respectively. We restate the following reduction:

\begin{lemma}[\cref{LemReduction}, Restate]\label{LemReduction2}
    For $P=S(T+1)$, we have
    $$\delta(S,T)\le 2\nu(P,T).$$
\end{lemma}

We briefly demonstrate the proof idea and highlight the difference with \cite{qipeng:2023} below.
\paragraph{Setup.}  
Let $\pi \leftarrow S_N$ be a random permutation. An $(S,T)$-auxiliary algorithm $\cA$ for the permutation inversion problem is specified by an $S$-qubit advice $\ket{\sigma_{\pi}}$ and a unitary $U^{\pi}_y$ which can be computed with $T$-quantum queries to $\pi$. The unitary $U^{\pi}_y$ operates on the register $\bA=\bS\otimes\bX\otimes\bL$, initialized as $\ket{\sigma_\pi}_{\bS}\ket{0}_{\bX,\bL}$ and aims to produce $\pi^{-1}(y)\in[N]$ in the register $\bX$. We call the algorithm $\cA$ \emph{uniform} if $\ket{\sigma_\pi}=\ket{0^S}$ for all $\pi$.

As in \cite{qipeng:2023}, define the POVMs:
$$P^{\pi}_y:=(U^{\pi}_y)^{\dagger} V^{\pi}_y U^{\pi}_y$$
and its average $P^{\pi}:=\E[y]{P^{\pi}_y}$ over challenges $y\in [N]$ to determine the correctness of an answer computed by the algorithm specified above where
$$V^{\pi}_y:=I_{\bS}\otimes\ket{\pi^{-1}(y)}\bra{\pi^{-1}(y)}_{\bX}\otimes I_{\bL}$$
is the projection for verifying the answer. The idea is to decompose $P^\pi=\sum_i p_{\pi,i}\ket{\phi_{\pi,i}}\bra{\phi_{\pi,i}}$ in the eigenbasis and write the initial state $\ket{\sigma_\pi}_{\bS}\ket{0}_{\bX,\bL}=\sum_i \alpha_{\pi,i}\ket{\phi_{\pi,i}}$ and analyze the success probability $\delta(S,T)$ as a quantity that depends only on the eigenvalues $\{p_{\pi,i}\}$ and coefficients $\{\alpha_{\pi,i}\}$ (see (\ref{EqEpsilon}) below).

\paragraph{Alternating measurement game.}  
The $g$-alternating measurement game is defined exactly as in \cite[Definition 6.3]{qipeng:2023}, alternating between measuring whether the algorithm succeeds on a fresh challenge by initializing a challenger register with a uniform superposition of challenges $y$ and applying the joint register with an algorithm by $\sum_{y\in [N]}\ket{y}\bra{y}\otimes P^{\pi}_y$ and rewinding the challenge register to a uniform superposition for $g$-rounds.

Notice that an alternating measurement game is associated with an algorithm $\cA$ with $S$-quantum advice and $T$-quantum queries and we denote $\epsilon_{\cA}^{\otimes g}(S,T)$ be the success probability of the $g$-alternating measurement game associated with $\cA$. Let $\epsilon^{\otimes g}(S,T)$ and $\epsilon^{\otimes g}(T)$ be the maximum success probability of alternating measurement game associated to all such $\cA$ and all \emph{uniform} $\cA$, respectively. For $g=1$, this coincides with the original algorithm $\cA$ and we have $\epsilon^{\otimes 1}(S,T)=\delta(S,T)$.

\paragraph{Proof sketch.}  
The analysis follows the three-step outline of \cite{qipeng:2023}:
\begin{enumerate}[label=(\roman*)]
    \item $(\delta(S,T))^g=(\epsilon^{\otimes 1}(S,T))^g \le \epsilon^{\otimes g}(S,T)$;
    \item $\epsilon^{\otimes g}(S,T)\le 2^S \epsilon^{\otimes g}(T)$;
    \item $\epsilon^{\otimes g}(T)\le \nu(P,T)^g$ with $P=g(T+T_{\Samp}+T_{\Verify})$.
\end{enumerate}
The $T_{\Samp}$ and $T_{\Verify}$ above are the numbers of queries needed to sample a challenge and verify an answer, so $T_{\Samp}=0$ and $T_{\Verify}=1$ for our case.
By setting $g=S$, we can immediately obtain Lemma \ref{LemReduction2}.

The key point is that the formula in \cite[Theorem 6.4]{qipeng:2023} becomes
\begin{equation}\label{EqEpsilon}
    \epsilon^{\otimes g}_{\cA}(S,T)
    = \tfrac{1}{N!}\sum_{\pi\in S_N}\sum_i |\alpha_{\pi,i}|^2 \cdot p_{\pi,i}^g
\end{equation}
in the random permutation model as it is obtained for each fixed $\pi$ before averaging, and thus does not depend on the oracle distribution.

The inequalities (i) is obtained by applying Jensen's inequality on the formula (\ref{EqEpsilon}), essentially using the convexity of $x^g$. For (ii), we can reduce any algorithm with quantum advice to uniform algorithm by guessing the advice. More precisely, we can always prepare the maximally mixed state on the oracle register with a loss of success probability up to $2^{-S}$. The inequality (iii) is more involved. For any adversary $\cA$ described above,
$$\epsilon_{\cA}^{\otimes g}(S,T)=\prod_{i=1}^g \P[]{b_i=0\mid b_1=\cdots=b_{i-1}=0}\le \P[]{b_g=0\mid b_1=\cdots=b_{g-1}=0}^g$$
where $b_i$ is a bit determining whether $\cA$ wins for the $i$-th round of the $g$-alternating game. The last inequality uses the observation that
$$\P[]{b_i=0\mid b_1=\cdots=b_{i-1}=0}=\frac{\epsilon_{\cA}^{\otimes i}(S,T)}{\epsilon_{\cA}^{\otimes i-1}(S,T)}$$
is monotonically increasing on $i$ again using the formula (\ref{EqEpsilon}) and the convexity (see \cite[Lemma 3.2]{qipeng:2023}). Finally, the conditional probability $\P[]{b_g=0\mid b_1=\cdots=b_{g-1}=0}$ can be reformulated as the success probability of an adversary in $P$-bit-fixing model (see \cite[Figure 6]{qipeng:2023} for this reduction) and we can get the desired bound.

\section{Representation Theory of Symmetric Group}\label{AppRep}
In this appendix, we collect the background on the representation theory of symmetric groups needed in this work. The material is standard: general references include \cite{serre:1977} for the theory of finite groups and \cite{jk:1984} for the case of symmetric groups. For connections to quantum query complexity, see also \cite{rosmanis:2014}.

Section \ref{AppSecRep} provides the general definitions and fundamental results of representation theory. Section \ref{AppSecSym} specializes to symmetric groups, emphasizing the correspondence with Young diagrams and the associated combinatorics. Section \ref{AppSecReg} examines the regular representation, which serves as a key example in our applications.

\subsection{Representation Theory of Finite Groups}\label{AppSecRep}
Representation theory may be viewed as linear algebra enhanced with the additional structure of a group action. In fact, ordinary linear algebra can be recovered as the representation theory of the trivial group.

Throughout we consider only finite groups acting on finite-dimensional complex Hilbert spaces, unless explicitly stated. We denote groups abstractly by $G,H$ and Hilbert spaces by calligraphic symbols such as $\cH,\cF$.

\begin{definition}[Representation]\label{DefRep}
    A \emph{representation} of a group $G$ is a pair $\rho=(\cH,V)$ where $\cH$ is a finite-dimensional Hilbert space and
    \[G\rightarrow \U(\cH)\qquad g\mapsto V_g\]
    is a homomorphism into the unitary group of $\cH$. Equivalently, $V_g V_h = V_{gh}$ for all $g,h\in G$ and $V_e=I$. The dimension of $\rho$ is $\dim \cH$, denoted by $d_{\rho}$
\end{definition}
\begin{remark}
In much of the literature, $\cH$ is only a vector space and $V_g\in\GL(\cH)$ need not be unitary. Since every representation is equivalent to a unitary one, we restrict to unitary representations, which is natural in the context of quantum computation.
\end{remark}

\begin{definition}[Homomorphism and Isomorphism]\label{DefGlinear}
    Let $\rho=(\cH,V)$ and $\rho'=(\cH',V')$ be $G$-representations. A linear map $L:\cH\to\cH'$ is a \emph{$G$-homomorphism} if
    $$L\circ V_g=V'_g\circ L$$
    for all $g\in G$. It is an \emph{isomorphism} if $L$ is invertible. In this case we write $\cH\simeq \cH'$.
\end{definition}

\begin{definition}[Subrepresentation]\label{DefSubrep}
    A subspace $\cH'\subseteq \cH$ is a \emph{subrepresentation} if it is invariant under the action of $G$:
    $$V_g(\cH')\subseteq \cH'$$
    for all $g\in G$. In particular $\cH'$ is itself a representation of $G$.
\end{definition}

Every representation has the trivial subrepresentations $0$ and $\cH$. The fundamental building blocks are the following:

\begin{definition}[Irreducible Representations]\label{DefIrrep}
A $G$-representation $\cH$ is \emph{irreducible} (or an \emph{irrep}) if it has no nontrivial subrepresentation. Otherwise it is \emph{reducible}. The set of all isomorphism classes of irreps of $G$ is denoted $\widehat G$. When convenient, we write $\widehat G = \{\rho_1,\dots,\rho_r\}$ with one representative of each isomorphism class.
\end{definition}

For example, if $G$ is trivial, $\widehat G$ contains only the one-dimensional representation. For symmetric groups $S_N$, however, $\widehat{S_N}$ admits a rich combinatorial description via Young diagrams.

A key tool is Schur’s Lemma:
\begin{fact}[Schur Lemma]\label{FactSchur}
    If $h:\cH\to\cH'$ is a $G$-homomorphism between irreps, then
    \begin{itemize}
        \item if $\cH\simeq\cH'$, $h$ is scalar multiplication.
        \item otherwise, $h=0$.
    \end{itemize}
\end{fact}

\paragraph{Operators on Representations.} Just as in linear algebra, we can combine representations in several ways.

\begin{definition}[Direct Sum]\label{DefDirect}
    If $\rho=(\cH,V)$ and $\rho'=(\cH',V')$ are $G$-representations, their \emph{direct sum} is
    $$\rho\oplus \rho':=(\cH\oplus \cH',V\oplus V'),\qquad g\mapsto
    \begin{pmatrix} 
V_g &   \\ 
 & V'_g
\end{pmatrix}.$$
\end{definition}

\begin{definition}[Tensor Product]\label{DefTensor}
    If $\rho=(\cH,V)$ is a $G$-representation and $\rho'=(\cH',V')$ an $H$-representation, their \emph{tensor product} is a $(G\times H)$-representation
    $$\rho\otimes\rho':=(\cH\otimes \cH',V\otimes V'),\qquad (g,h)\mapsto V_g\otimes V'_h.$$
\end{definition}

\begin{fact}[Irreps of a Product Group]\label{FactProduct}
If $\widehat G=\{\rho_1,\dots,\rho_r\}$ and $\widehat H=\{\sigma_1,\dots,\sigma_s\}$, then
$$\widehat{G\times H}=\{\rho_i\otimes \sigma_j\mid 1\le i\le r, 1\le j\le s\}.$$
\end{fact}

\begin{definition}[Restriction]\label{DefRestrict}
    If $\rho=(\cH,V)$ is a $G$-representation and $H\subseteq G$ a subgroup, the \emph{restriction} is
    $$\Res[H][G] \rho:=(\cH,V|_{H}),\qquad h\mapsto V_h.$$
\end{definition}

\paragraph{Decomposition of Representations.}
\begin{fact}[Orthogonal Complements]\label{FactOrth}
    If $\cH'\subseteq\cH$ is a $G$-subrepresentation, then its orthogonal complement $(\cH')^\perp$ is also a $G$-subrepresentation and $\cH=\cH'\oplus (\cH')^{\perp}$.
\end{fact}

\begin{fact}[Unique Decomposition]\label{FactIrrepDecomp}
    If $\widehat G={\rho_1,\dots,\rho_r}$, then any $G$-representation $\cH$ decomposes (up to isomorphism) uniquely as
    \begin{equation}\label{EqUniquDecomp}
        \cH\simeq \bigoplus_{i=1}^r \cH_i^{m_i}
    \end{equation}
    for uniquely determined multiplicities $m_i\ge 0$.
\end{fact}
The direct-sum decomposition (\ref{EqUniquDecomp}) is not canonical as $\cH_i$ is not a subspace of $\cH$. A more intrinsic version uses isotypic subspaces:

\begin{definition}[Isotypic Subspace]\label{DefIsotypic}
      For a representation $\cH$ and an irrep $\rho\in\widehat G$, the \emph{$\rho$-isotypic subspace} $\cH_\rho$ is the subspace generated by all subrepresentations of $\cH$ isomorphic to $\rho$.
\end{definition}

\begin{fact}[Isotypic Decomposition]\label{FactIsotyDecomp}
    Any representation decomposes canonically as
    $$\cH=\bigoplus_{i=1}^r \cH_{\rho_i}$$
    and the isotypic subspaces are pairwise orthogonal subrepresentation of $\cH$.
\end{fact}

\subsection{Representation Theory of Symmetric Groups}\label{AppSecSym}
The representation theory of symmetric groups is closely tied to combinatorics through the correspondence with \emph{Young diagrams}. This connection provides a powerful framework for understanding and computing irreducible representations.  

A symmetric group
\[
S_{[N]}:=\big\{\pi:[N]\rightarrow [N]\;\mid\;\pi \text{ is a bijection}\big\}
\]
consists of all permutations of $N$ elements. We also write $S_N$ when the underlying set is clear.

\begin{definition}[Young Diagram]\label{DefYoung}
    For $N \geq 0$, a \emph{Young diagram} $\lambda$ of size $N$ is a partition of $N$, \emph{i.e.}, a sequence of positive integers $\lambda=(\lambda_1,\ldots,\lambda_r)$ with $\lambda_1\ge\dots\ge \lambda_r$. For $N=0$, we set $\rho=\emptyset$.  

    A Young diagram can be visualized as boxes arranged in left-aligned rows of lengths $\lambda_1,\dots,\lambda_r$ in weakly decreasing order. We write $\rho \vdash N$ if $\rho$ has size $N$.  

    The \emph{transpose} of $\lambda$ is $\lambda^{\perp}=(\lambda^{\perp}_1,\dots,\lambda^{\perp}_c)$, where $\lambda^{\perp}_j:=\max\{i\mid \lambda_i\ge j\}$ and $c:=\lambda_1$.
\end{definition}

\begin{definition}
    A box $(i,j)$ is in a Young diagram $\lambda=(\lambda_1,\dots,\lambda_r)$ if $\lambda_i\ge j$. This corresponds to the box in row $i$ and column $j$.
\end{definition}

\begin{figure}[ht]
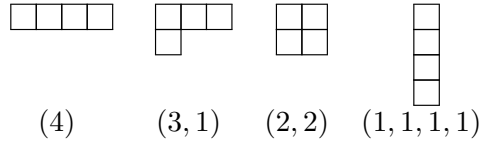

\centering
\begin{tabular}{ c c c c }

\ytableausetup{smalltableaux}
\begin{ytableau}
{} & {} & {} & {}
\end{ytableau} &

\ytableausetup{smalltableaux}
\begin{ytableau}
{} & {} & {} \\
{}
\end{ytableau} &

\ytableausetup{smalltableaux}
\begin{ytableau}
{} & {} \\
{} & {} 
\end{ytableau} &

\ytableausetup{smalltableaux}
\begin{ytableau}
{} \\
{} \\
{} \\
{}
\end{ytableau} 
 \\ 
$(4)$ & $(3,1)$ & $(2,2)$ & $(1,1,1,1)$
\end{tabular}
\caption{The Young diagrams (top) and corresponding partitions (bottom) of size $4$.}
\end{figure}

The central fact in the representation theory of symmetric groups is the following:

\begin{fact}[Specht Module]\label{FactCorres}
For any $\lambda\vdash N$, there exists an irreducible representation $\mathcal{S}_\lambda$ of $S_N$, called the \emph{Specht module}, such that
\[
\widehat{S}_N=\{\mathcal{S}_\lambda \mid \lambda\vdash N\}.
\]
That is, every irrep of $S_N$ is isomorphic to $\mathcal{S}_\lambda$ for some Young diagram $\lambda$.
\end{fact}

The explicit construction of $\mathcal{S}_\lambda$ is not needed here; we only rely on its properties. By abuse of notation, we use $\lambda$ to denote both a Young diagram and its corresponding irrep.  
\begin{definition}[Hook length]\label{HookLength}
   Given $\lambda\vdash N$ and $(i,j)\in \lambda$, the \emph{hook length} is
\[
h_{\lambda}(i,j):=(\lambda_i-j)+(\lambda^{\perp}_j-i)+1.
\]
   Define
\[
h(\lambda):=\prod_{(i,j)\in\lambda} h_{\lambda}(i,j).
\]
\end{definition}

\begin{fact}[Hook length formula]\label{FactFormulaHook}
     For a Young diagram $\lambda \vdash N$, the dimension $d_\lambda$ of the corresponding irrep is
\[
d_\lambda=\frac{N!}{h(\lambda)}.
\]
\end{fact}

\begin{figure}[ht]
\begin{equation*}
\vcenter{
\hbox{
\begin{ytableau}
{7} & {6} & {4} & {2} & {1} \\
{4} & {3} & {1} \\
{2} & {1} 
\end{ytableau} }}
\end{equation*}
\caption{Hook lengths for $\lambda=(5,3,2)$. The dimension is $d_{\lambda}=\tfrac{10!}{7\cdot 6\cdot 4\cdot 4\cdot 3\cdot 2\cdot 2}=300$.}
\end{figure}

When restricting a representation to a subgroup, irreps typically decompose into smaller irreps. For symmetric groups this decomposition is fully described by the following:

\begin{definition}
    Let $\lambda \vdash N$ and $\mu \vdash (N-1)$. We write $\mu\prec\lambda$ if $\mu$ is obtained from $\lambda$ by removing a box.
\end{definition}

For $y\in [N]$, let
\[
S_{[N]\setminus\{y\}}:=\{\pi\in S_{[N]} \mid \pi(y)=y\}
\]
be the subgroup of permutations fixing $y$. For $\mu\vdash N-1$, we write $\mu_y$ for the corresponding irrep of $S_{[N]\setminus\{y\}}$ (the subscript indicates the fixed point).
\begin{fact}[Branching Rule]\label{FactBranch}
    For any $\lambda\vdash N$ and $y\in[N]$,
    $$\Res[S_{[N]\setminus\{y\}}][S_{[N]}] \lambda\;\simeq\; \bigoplus_{\mu\prec\lambda}\mu_y$$
    where $\lambda$ and $\mu_y$ are irreps of $S_{[N]}$ and $S_{[N]\setminus\{y\}}$, respectively.
\end{fact}
\begin{figure}[ht]\label{FigBranch}
\centering
\begin{equation*}
\Res[S_{9}][S_{10}]\left(
\vcenter{
\hbox{
\begin{ytableau}
\ & \ & \ & \ & \ & \ \\
\ & \ & \ \\
\
\end{ytableau}
}
}
\right)=
\vcenter{
\hbox{
\begin{ytableau}
\ & \ & \ & \ & \ \\
\ & \ & \ \\
\
\end{ytableau}
}
}\oplus
\vcenter{
\hbox{
\begin{ytableau}
\ & \ & \ & \ & \ & \ \\
\ & \ \\
\
\end{ytableau}
}
}\oplus
\vcenter{
\hbox{
\begin{ytableau}
\ & \ & \ & \ & \ & \ \\
\ & \ & \ 
\end{ytableau}
}
}
\end{equation*}

\caption{Restriction of the $S_{10}$-representation $\lambda=(6,3,1)$ to $S_9$. The fixed point $y\in[10]$ is unspecified.}
\end{figure}

We also introduce a convenient non-standard notation for our applications.

\begin{definition}\label{DefBarTheta}
For a Young diagram $\theta$ of size $k<N$, define
\[
\overline{\theta}:=(N-k,\theta), \quad
\overline{\theta}_*:=(N-k-1,\theta),
\]
whenever these are valid Young diagrams.
\end{definition}
An illustration of Definition \ref{DefBarTheta} is given in Figure \ref{FigBar} in Section \ref{SecDecomp}. The following lemma, due to \cite[Claim~7]{rosmanis:2014}, will be used in proving Lemma~\ref{LemAvg}. For completeness we include its proof via the hook-length formula.

\IneqDim*

\proof
Let $\zeta(i)$ be the number of boxes below $(i,1)$ for $i\in [N-k]$ (so $\zeta(i)=0$ if $i>k$). By Fact~\ref{FactFormulaHook},
\[
d_{\overline{\theta}}
=\frac{N!}{h_{\overline{\theta}}}
=\frac{N!}{h_{\theta}\cdot\prod_{i=1}^k (N+1-k+\zeta(i))\cdot (N-2k)!},
\]
and
\[
d_{\overline{\theta}_*}
=\frac{(N-1)!}{h_{\overline{\theta}_*}}
=\frac{(N-1)!}{h_{\theta}\cdot\prod_{i=1}^k (N-k+\zeta(i))\cdot (N-2k-1)!}.
\]
Hence
\[
\frac{d_{\overline{\theta}_*}}{d_{\overline{\theta}}}
=\frac{N-2k}{N}\prod_{i=1}^k \frac{N+1-k+\zeta(i)}{N-k+\zeta(i)}
\;\;\ge\;\; \frac{N-2k}{N}.
\]
\qed

\subsection{Regular Representation of Symmetric Group}\label{AppSecReg}
We now examine the \emph{regular representation} of the symmetric group, which is universal in the sense that it contains every irrep of $S_{[N]}$ as a component.  

Let
\[
\mathcal{F}=\Span\{\ket{\pi}\mid \pi\in S_{[N]}\}
\]
be the Hilbert space spanned by basis vectors indexed by permutations. We define an $S_{[N]}\times S_{[N]}$-representation
\[
V:S_{[N]}\times S_{[N]}\to \mathsf{U}(\mathcal{F}), \qquad (\pi_{D},\pi_{R})\mapsto V_{\pi_{D}}^{\pi_{R}},
\]
by
\[
V_{\pi_{D}}^{\pi_{R}}\ket{\pi}:=\ket{\pi_{R}\circ \pi\circ \pi_{D}^{-1}},
\]
extended linearly. This is called the \emph{regular representation} of $S_{[N]}$.

Restricting this action gives two natural $S_{[N]}$-representations:
\begin{itemize}
    \item the \emph{left} action:
    \[
    \pi_{D}\in S_{[N]}\mapsto V_{\pi_{D}},\qquad V_{\pi_{D}}\ket{\pi}=\ket{\pi\circ \pi_{D}^{-1}},
    \]
    \item the \emph{right} action:
    \[
    \pi_{R}\in S_{[N]}\mapsto V^{\pi_{R}},\qquad V^{\pi_{R}}\ket{\pi}=\ket{\pi_{R}\circ \pi}.
    \]
\end{itemize}

By Fact~\ref{FactProduct} and Fact~\ref{FactCorres}, every irrep of $S_{[N]}\times S_{[N]}$ is of the form $\lambda\otimes \lambda'$, where $\lambda,\lambda'$ are Young diagrams of size $N$. We denote the isotypic subspace of $\lambda\otimes\lambda'$ in $\mathcal{F}$ by $\mathcal{H}_\lambda^{\lambda'}$, and use $\mathcal{H}_\lambda$ and $\mathcal{H}^\lambda$ for the left and right isotypic subspaces, respectively.

The regular representation has the following key property:

\begin{fact}\label{FactRegIsoty}
If $\lambda\neq \lambda'$, then $\mathcal{H}_\lambda^{\lambda'}=0$. Moreover,
\[
\mathcal{H}^{\lambda}_{\lambda}
=\mathcal{H}_{\lambda}
=\mathcal{H}^{\lambda}
\simeq \lambda\otimes \lambda.
\]
\end{fact}

Thus, we consistently denote this subspace by $\mathcal{H}_\lambda$. By Fact~\ref{FactIsotyDecomp}, we obtain the orthogonal decomposition
\[
\mathcal{F}=\bigoplus_{\lambda\vdash N}\mathcal{H}_\lambda.
\]

If $M:\mathcal{F}\to \mathcal{F}$ is a homomorphism of $S_{[N]}\times S_{[N]}$-representations, then by Schur’s lemma (Fact~\ref{FactSchur}), $M$ acts as a scalar $e_\lambda$ on each $\mathcal{H}_\lambda$, and vanishes between distinct subspaces. Hence
\begin{equation*}\label{EqMApp}
M=\sum_{\lambda\vdash N} e_\lambda \Pi_\lambda,
\end{equation*}
where $\Pi_\lambda$ is the orthogonal projection onto $\mathcal{H}_\lambda$.  
Note that this decomposition need not hold if $M$ is only a homomorphism of the left or right $S_{[N]}$-representation.

\end{document}